\documentclass[amssymb,graphics,floatfix, nobibnotes,prd,superscriptaddress,nofootinbib,tightenlines,nobibnotes, aps, 12pt]{revtex4-2}
\pdfoutput=1

\usepackage{iftex}
\ifPDFTeX
  \usepackage[utf8]{inputenc}
  \usepackage[T1]{fontenc}
\else
  \usepackage{fontspec}
\fi

\usepackage{pst-all}
\usepackage{multirow}
\usepackage{wrapfig,lipsum}
\usepackage{booktabs}
\usepackage{listings}
\usepackage{lineno} 				
\usepackage{amsmath}
\usepackage{natbib}
\usepackage{txfonts}
\usepackage{comment}
\usepackage{graphics,graphicx}
\usepackage{amssymb,bm,cancel}
\usepackage{ulem}

\newcommand{\sfb}[1]{{\bm{{\sf #1}}}}

\begin{document}
\title{The Design Strain Sensitivity of the Schenberg Spherical Resonant Antenna for Gravitational Waves}
\author{Liccardo V.}
\email{vic2000@hotmail.it}
\affiliation{Instituto Nacional de Pesquisas Espaciais, 12227-010 São José dos Campos, São Paulo, Brazil}
\author{Lenzi C. H.}
\email{chlenzi1980@gmail.com}
\affiliation{Instituto Tecnológico de Aeronáutica, 12228-900 São José dos Campos, São Paulo, Brazil}
\author{Marinho Jr. R. M.}
\email{marinho.rubens@gmail.com}
\affiliation{Instituto Tecnológico de Aeronáutica, 12228-900 São José dos Campos, São Paulo, Brazil}
\author{Aguiar O. D.}
\email{odylio.aguiar@inpe.br}
\affiliation{Instituto Nacional de Pesquisas Espaciais, 12227-010 São José dos Campos, São Paulo, Brazil}
\author{Frajuca C.}
\email{Frajuca@gmail.com}
\affiliation{Universidade Federal do Rio Grande, 96203-900 Rio Grande, Rio Grande do Sul, Brazil}
\author{Bortoli F. S.}
\email{bortoli@ifsp.edu.br}
\affiliation{Instituto Federal de São Paulo, 01109-010 São Paulo, São Paulo, Brazil}
\author{Costa C. A.}
\email{cescosta@gmail.com}
\affiliation{Instituto Nacional de Pesquisas Espaciais, 12227-010 São José dos Campos, São Paulo, Brazil}
\date{\today}

\begin{abstract}
The main purpose of this study is to review the Schenberg resonant antenna transfer function and to recalculate the antenna design strain sensitivity for gravitational waves. We consider the spherical antenna  with six transducers in the semi dodecahedral  configuration. When coupled to the antenna, the transducer-sphere system will work as a mass-spring system with three masses. The first one is the antenna effective mass for each quadrupole mode, the second one is the mass of the mechanical structure of the transducer first mechanical mode and the third one is the effective mass of the transducer membrane that makes one of the transducer microwave cavity walls. All the calculations are done for the degenerate (all the sphere quadrupole mode frequencies equal) and non-degenerate sphere cases. We have come to the conclusion that the ``ultimate'' sensitivity of an advanced version of Schenberg antenna (aSchenberg) is around the standard quantum limit (although the parametric transducers used could, in principle, surpass this limit). However, this sensitivity, in the frequency range where Schenberg operates, has already been achieved by the two aLIGOs in the O3 run, therefore, the only reasonable justification for remounting the Schenberg antenna and trying to place it in the sensitivity of the standard quantum limit would be to detect gravitational waves with another physical principle, different from the one used by laser interferometers. This other physical principle would be the absorption of the gravitational wave energy by a resonant mass like Schenberg. 
\end{abstract}

\maketitle

\section{Introduction}\label{sec-introd}

Gravitational waves (GW) are ripples in the fabric of space-time generated by the acceleration of massive cosmic objects. These ripples move at the speed of light and can excite quadrupolar normal-modes of elastic bodies. The first detection of GWs from the inward spiral and merger of a pair of Black Holes (BH) (GW150914) has been widely discussed in the literature \cite{Abbott16, Abbott16b, Abbott16c, Abbott16d}. Furthermore, the recent simultaneous detection of the electromagnetic counterpart with GWs from a binary Neutron Star (NS) merger (GW170817) has officially begun the era of multi-messenger astronomy involving GWs \cite{Abbott17, Abbott17b}. Studying the universe with these two fundamentally different types of information will offer the possibility of a richer understanding of the astrophysical scenarios as well as of nuclear processes and nucleosynthesis. For the first time in the GW astronomy, it has been possible to determine the position in the sky of the source thanks to the detection, at the same time, of the three interferometers of the LIGO/Virgo collaboration \cite{Abbott17}. 

\begin{figure}[!ht]
\begin{center}
   \includegraphics[scale=0.55]{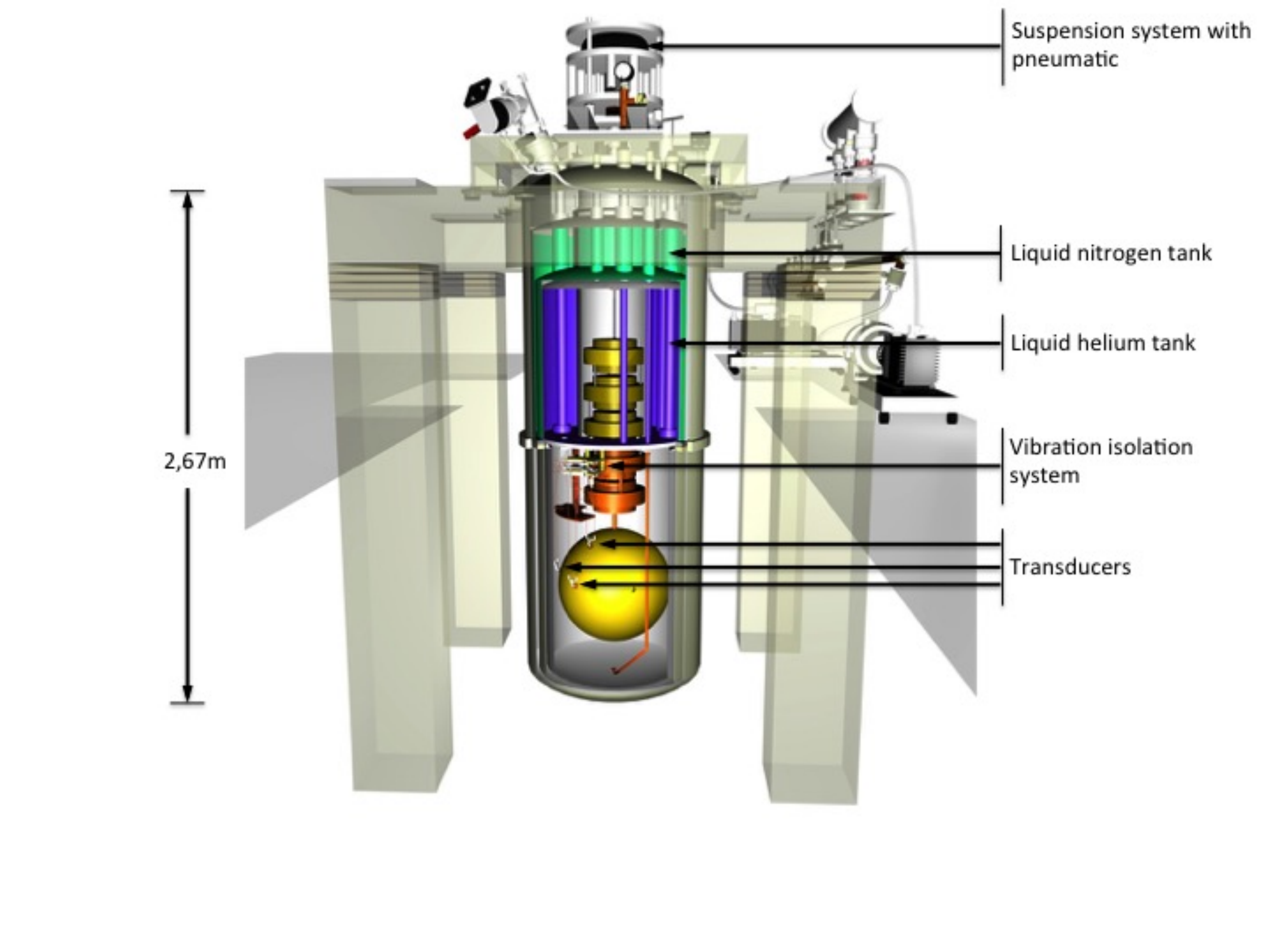}
     \caption{The Schenberg antenna where nine parametric transducers monitor the fundamental modes of vibration of the resonant spherical mass (credit: Xavier P. M. Gratens).}
     \label{detec}
     \end{center}
\end{figure}

The Mario Schenberg Brazilian detector is based on the detection of five quadrupole modes 
relative to the mechanical vibrations of a spherical resonant-mass of 
$M_S= 1124$ kg and radius $R=32.33$ cm (Fig. \ref{detec}). The operating frequency band is 3.15 - 3.26 kHz. The antenna is made of a CuAl(6\%) alloy, 
which has a high mechanical quality factor Q $\sim$ $2\times10^{6}$ at 4 K. The system is suspended by a 
vibration isolation system, capable of attenuating external vibrations by about 300 dB \cite{Melo02, suspensionBortoli}. 
The instrument will be maintained at low temperatures ($\sim$ 4 K) by cryogenic
chambers (dewars), cooled down by a He flow  \cite{Waard02}. The antenna is coupled to parametric 
transducers that will monitor the vibrations of the quadrupolar/monopolar normal modes of the sphere 
\cite{Paula15, Liccardo16, Cabling, Quadrupole, dilutionrefrigerator}.  One of the main advantages of a GW spherical resonant antenna is its omnidirectional sensitivity, which makes it equally responsive to all wave directions and polarizations. 
Spherical resonant-mass antennas have been already intensively studied \cite{Coccia95, Harry96, Lobo00}. 
The designed antenna transduction system consists of nine transducers fixed on the surface of the sphere, 
six of which follow the truncated icosahedron configuration proposed by Johnson and Merkowitz \cite{Johnson93}. 
This configuration presents some benefits and allows the simplification of the equations of motion, 
the determination of the GW direction in the sky, and facilitates the interpretation of the signal. 
For more details on the Schenberg antenna, the reader is referred to \cite{Aguiar06, Aguiar12} and references therein. It is important to mention that, in addition to being a device to try to detect gravitational waves, the Schenberg antenna could also be used to test the hypothesis that the ripples in the curvature of the fabric of space-time can be scaled by a more minute “action”, whose detection requires sensitivities beyond the standard quantum limit \cite{Messina15}. 
On the other hand, the Schenberg detector can also be used to test alternative theories of gravitation, such as the reference \cite{Paula_2004} which, having a massive graviton, has six polarization states.

The plan of the paper is as follows: in Sec.(\ref{sec:NS-BH}), 
we consider the emission of GWs from the spiraling of a NS-BH binary system and we discuss the detectability of this system by the Schenberg antenna. 
Then, we discuss the interaction of GWs with matter in Sec.(\ref{gwsection}).
The detector model is introduced in Sec.(\ref{sec:ITM}), which is followed by the calculation of the response function of the antenna. Final considerations as well as the discussion of the results are presented in Sec.(\ref{sec:disc}).

\section{Gravitational Waves from NS-BH binary systems}
\label{sec:NS-BH}

Coalescence of NS-BH binaries is one of the most promising GW sources for ground-based antennas. NS-BH systems are believed to be formed as a result of two supernovae in a massive binary system \cite{Lee10, Shibata11}. GWs from binaries involving NS represent a tool to study NS properties like the radius, compactness, and tidal deformability. Knowledge of NS properties will allow constraining the equation of state of nuclear-density matter \cite{Foucart14},  giving us valuable information on nuclear physics. After the formation of the system, the orbital separation decreases gradually due to the long-term gravitational radiation reaction (i.e., two objects are in an adiabatic inspiral motion), and eventually, the two objects merge into a BH. The final fate of the binary depends primarily on the mass of the BH and the compactness of the NS. However, a detailed analysis has shown that the BH spin and the NS equation of state also play an important role in determining the final fate \cite{Shibata11}. The effective-one-body (EOB) formalism was introduced \cite{Buonanno99, Buonanno00} as a promising approach to describe analytically the inspiral, merger, and ringdown waveforms emitted during a binary merger. 
Among the candidates of electromagnetic counterparts, a short-hard Gamma-Ray Burst (GRB) and its afterglow are vigorously studied both theoretically and observationally \cite{Nakar07, Berger14}.
For a deeper analysis of NS-BH binaries, see \cite{Shibata11}.

In this section, we discuss the GW signal produced by the coalescence of a 
non-spinning 1.4 -- 3.0 $M_{\odot}$ NS-BH binary system, disregarding finite-size effects such as tidal deformation. 
The narrow frequency window of the antenna constrains the BH mass to be $\lesssim$ 3 $M_{\odot}$. Compact binary systems emit periodic GWs, whose frequencies sweep the spectrum until they reach their maximum values when they are close to the coalescence. The characteristic amplitude and the frequency of GWs near the last orbit are given by \cite{Shibata11}

\begin{equation}\label{signal1}
h \approx 3.6 \times {10^{ - 22}}\left( {\frac{{{M_{BH}}}}{{6{M_ \odot }}}} \right)\left( {\frac{{{M_{NS}}}}{{1.4{M_ \odot }}}} \right)\left( {\frac{{6GM}}{{{c^2}r}}} \right)\left( {\frac{{0.1{\rm{ Gpc}}}}{D}} \right),
\end{equation}

\begin{equation}\label{signal2}
f \approx \frac{\omega }{\pi } \approx 594{\rm{Hz}}{\left( {\frac{{6GM}}{{{c^2}r}}} \right)^{\frac{3}{2}}}\left( {\frac{{7.4{M_ \odot }}}{M}} \right),
\end{equation}

where $\omega$ is the angular velocity, $M$ = $M_{BH}$ + $M_{NS}$, and $r$ and $D$ are the orbital separation and the distance to the source, respectively. The binary system studied may be in principle detected since the frequency of the gravitational signal $\sim$ 1 ms before coalescing falls in the band of the Brazilian antenna.
NS-BH mergers are also potential targets of interferometers GW detectors. Since these kinds of antennas are sensitive in a much broader frequency range ($\sim$ 10 - 4000 Hz) they will detect the signal before the Schenberg 
antenna (during the inspiral phase). It is worth noting that due to the truncated icosahedron configuration the antenna is able to determine the polarization and the position of astrophysical sources of the GW \cite{Magalhaes95,Magalh97,Lenzi1,Lenzi2}.

\begin{figure}[!ht]
\begin{center}
   \includegraphics[scale=0.8]{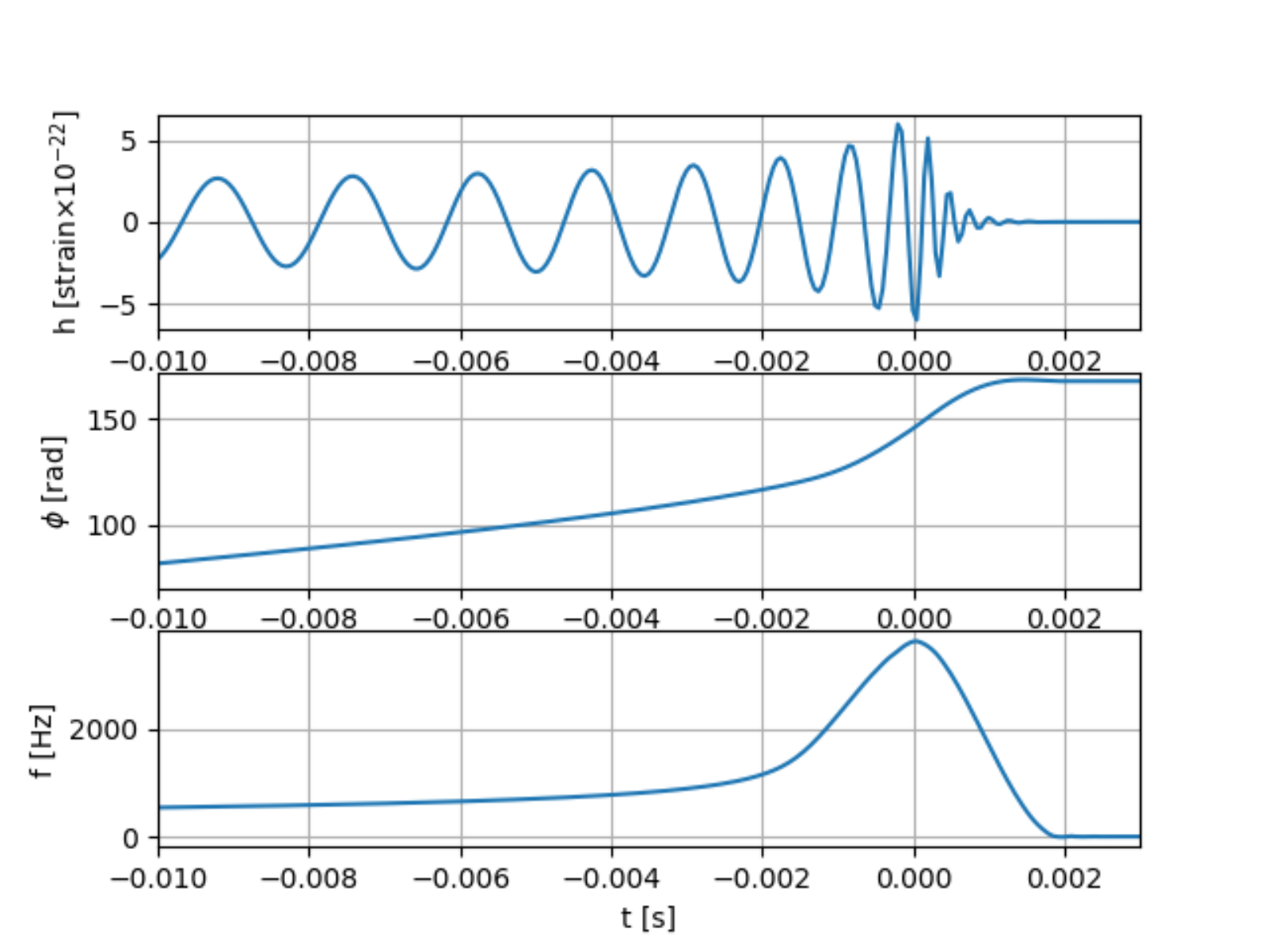}
     \caption{The GW strain signal produced by the coalescence of a non-spinning 1.4 -- 3.0 $M_{\odot}$ NS-BH binary system ($Top$), phase ($Middle$) and frequency ($Bottom$) are plotted as function of the time before merging.}
     \label{evol}
     \end{center}
\end{figure}

There are a large number of waveform families in the literature, obtained from considerations about the type of source and approximation procedures used for the simulation (numerical relativity (NR), EOB formalism, post-Newtonian (PN) approximation, etc.). The gravitational signal for our analysis was generated using the PyCBC software package \cite{Canton14, Usman15}. The waveform employed is one of those that are used by LIGO/Virgo, that is, the effective-one-body model tuned to numerical relativity (EOBNRv2). PN results are good as long as the velocities of the objects are not extreme relativistic. However, as the two objects orbit around each other, they lose energy through the emission of GWs, and their distance shrinks along with an increase in velocity. Consequently, PN predictions become more and more inaccurate the closer the binary gets to the merger, while the EOB approach, close to the merger, provides better accuracy by calibrating higher-order vacuum terms to NR waveforms. The EOBNRv2 waveform is believed to be sufficiently accurate to search for signals from non-spinning coalescing compact binaries in the aLIGO sensitive band.
The EOB formalism has been refined several times to incorporate additional information from NR. 
Depending on the number of available NR waveforms as well as the modifications introduced to the EOB description, various versions of such EOBNR models have been developed \cite{Damour09, Pan11}. It is beyond the scope of this paper to show the technical details of the EOB formalism and its extensions. Figure \ref{evol} also shows the waveform of the non-spinning NS-BH binary considered here. The waveform has also been re-sampled to be compatible with the sampling rate of the Schenberg antenna.


The coalescence rate of this type of system is very small and can be calculated indirectly. Upper limits ($\sim$ $10^{3}$ Gpc$^{-3}$ yr$^{-1}$) were given assuming that all short GRBs/kilonovae are linked with NS-BH mergers \cite{Nakar07} and from the assumption that all the r-process material were produced in NS-BH coalescences \cite{Bauswein14}.

There are indications that NS-BH binary has been directly observed  \cite{Abbott21} and an estimated rate density of $\sim$~0.04~$\times$~$10^{3}$~Gpc$^{-3}$~yr$^{-1}$ can also be derived from stellar evolution synthesis \cite{Abadie10, Dominik15}. In the present work, to evaluate the event rate related to NS-BH mergers, we follow Li $et$ $al$ \cite{Li17} and Abbott $et$ $al$ \cite{Abbott16a},
who constrain the merger rate to be less than 6500 Gpc$^{-3}$ yr$^{-1}$, assuming a population of binary systems of 1.4 -- 3 $M_{\odot}$. This estimate is sensitive to physical parameters, such as the equation of state of NS material and the mass/spin distribution of the BH. The upper limit of the rate decreases for BHs with larger masses. The expected rates for other transient sources are smaller and/or less reliable. In order to be detected, the amplitude of the GW signal needs to be compatible with the sensitivity of the antenna. 

For an advanced version of the Schenberg antenna (aSchenberg), which would operate around the standard quantum limit (Sec. \ref{sec:ITM}), gravitational signals with amplitude $h$ $\sim$ 10$^{-22}$ could be detected at the nominal frequency of the antenna.
In this case, a signal could be produced in GWs whose characteristic amplitude is $\sim$ 3 $\times$  10$^{-22}$ at distances of the order of 0.1 Gpc (Fig. \ref{evol}). In this volume, the event rate would be $\sim$ 3.6 yr$^{-1}$ at a SNR $\sim$ 1. 
This conclusion relies on the validity of the assumption that all observed kilonovae were associated with NS-BH coalescences.
In addition, many statistical studies based on the stellar evolution synthesis and supernova rates predict the rates at which NS-BH merge in the Milky Way and the nearby universe, assuming that Milky Way-like galaxies dominate, to be 1-10\% of that of NS-NS binaries (every $\sim$ $10^{6}$ - $10^{7}$ years) \cite{Voss03, Kalogera07, Kim08}. If we consider the contribution of elliptic galaxies the total coalescence rate of the universe could be increased by a significant fraction \cite{Kalogera10}.
These estimates show that the prospect for the detection of NS-BH mergers of 1.4 -- 3.0 $M_{\odot}$ by the Schenberg antenna can be very promising. 

\section{The interaction of GW with matter}
\label{gwsection}
As it is well known, a GW produces a tidal density force at time $t$ and at position $\bm x$
given by (sum over repeated indices implied)
\begin{equation}
    f_i^{GW}(\bm x,t)=\frac 12\rho \ddot h_{ij}(t)x_j,
    \label{fi1}
\end{equation}
where $\rho$ is the mass density and $\ddot h$ the second time derivative of the GW amplitude.
Since the Schenberg antenna has a resonant frequency about 3 kHz,
the wavelength of the GW detectable is about 100 km so we can use the value of $h_{ij}(t)$
at the center of the sphere. Eq.(\ref{fi1}) can be written in terms of the gradient of a potential

\begin{equation}
    \bm f^{GW}(\bm x,t)=-\bm\nabla\Phi(\bm x,t),
\end{equation}

where

\begin{equation}
    \Phi(\bm x,t)=
    -\frac 14\rho x_i\ddot h_{ij}(t)x_j=-\frac 14\rho r^2 n_i\ddot h_{ij}(t)n_j,
    \label{fi2}
\end{equation}

where $\bm n$ is the unit vector in the radial direction and $r$ the magnitude.
We can expand $\Phi(\bm x,t)$ in terms of the real spherical harmonics,
always used in this paper, $Y_{\ell m}^{\cal R}(\vartheta,\varphi)$, defined 
in terms of the traditional spherical harmonics

\begin{align}
    Y_{\ell,-m}^{\cal R}(\vartheta,\varphi)&=\sqrt{2}{\cal I}[Y_{\ell m}(\vartheta,\varphi)]\nonumber\\
    Y_{\ell 0}^{\cal R} &= Y_{\ell 0} \label{realY}\\
    Y_{\ell m}^{\cal R}(\vartheta,\varphi) &=\sqrt{2}{\cal R}[Y_{\ell m}(\vartheta,\varphi)]\nonumber.
\end{align}

The spherical harmonics obey the normalization condition

\begin{equation}
    \int_{\vartheta=0}^\pi\int_{\varphi=0}^{2\pi}
    Y_{\ell m}^{\cal R}Y_{\ell' m'}^{\cal R}
    \sin\vartheta d\vartheta d\varphi=\delta_{\ell\ell'}\delta_{mm'}.
\end{equation}

From now on we will omit the superscript $\cal R$ and write $Y_{\ell m}^{\cal R}=Y_{\ell m}$.
After the expansion we have (only terms with $\ell=2$, quadrupolar modes, survive)

\begin{equation}
    \Phi(\bm x, t) = -\sqrt{\frac{\pi}{15}}\rho r^2 \ddot h_m(t)Y_{2m},
    \label{phi}
\end{equation}

where the $h_m$ are the expansion coefficients so called spherical amplitudes given by

\begin{align}
    h_{-2} &=h_{12} \\\
    h_{-1} &=h_{23} \\
    h_0 &=\frac{\sqrt{3}}{2}h_{33} \\
    h_1 &=h_{13} \\
    h_2&=\frac 12(h_{11}-h_{22}).
\end{align}

The spherical amplitudes $h_m$ for a GW coming from the direction defined by the polar and azimuthal angles $(\theta,\phi)$ as seen from the lab frame 
is given by (see Appendix(\ref{hWFtohLF})):

\begin{align}
    h_{-2} &=\frac 12(1+\cos^2\theta)\sin2\phi h_++\cos\theta\cos2\phi h_\times\\
    h_{-1} &=-\frac 12\sin2\theta\sin\phi h_+-\sin\theta\cos\phi h_\times\\
    h_0 &=\frac{\sqrt{3}}{2}\sin^2\theta h_+\\
    h_1 &=-\frac 12\sin2\theta\cos\phi h_++\sin\theta\sin\phi h_\times\\
    h_2&=\frac 12(1+\cos^2\theta)\cos2\phi h_+-\cos\theta\sin2\phi h_\times.
\end{align}

In matrix notation and after making the rotation around the polarization angle $\psi$, we have 
\begin{equation}
    \begin{pmatrix}
    h_{-2}\\h_{-1}\\h_0\\h_1\\h_2
    \end{pmatrix}=
    \begin{pmatrix}
        \frac 12(1+\cos^2\theta)\sin2\phi & \cos\theta\cos2\phi\\
        -\frac 12\sin2\theta\sin\phi      & -\sin\theta\cos\phi\\
        \frac{\sqrt{3}}{2}\sin^2\theta    & 0\\
        -\frac 12\sin2\theta\cos\phi      & \sin\theta\sin\phi\\
        \frac 12(1+\cos^2\theta)\cos2\phi & -\cos\theta\sin2\phi
    \end{pmatrix}
    \begin{pmatrix}
        \cos2\psi & -\sin2\psi \\
        \sin2\psi & \cos2\psi
    \end{pmatrix}
    \begin{pmatrix}
    h_+\\
    h_\times
    \end{pmatrix}.
    \label{hs}
\end{equation}

Using Eq.(\ref{phi}) and the vector spherical harmonics (see Appendix(\ref{vsharmonics})) we obtain the expression of the
GW density force
\begin{equation}
    \bm f^{GW}=\sqrt{\frac{4\pi}{15}}\rho r\ddot h_m(t)
    \left(\bm Y_{2m}^L+\frac{\sqrt{6}}{2}\bm Y_{2m}^E\right).
    \label{fGW}
\end{equation}
In the case where $\bm f$ in the right hand side of Eq.(\ref{eqmovad}) is only of
GW origin, the overlap integral 
\begin{equation}
    f^{GW}_{n\ell m}=\int_V\bm\Psi_{n\ell m}(\bm x)\cdot\bm f^{GW}(\bm x,t)d^3x
\end{equation}
is the effective force on each mode of the sphere and
\begin{equation}
    \boldsymbol{\Psi}_{n\ell m}(\boldsymbol{x})=
    A_{n\ell}(r)\boldsymbol{Y}_{\ell m}^{L}(\theta,\phi)+
    B_{n\ell}(r)\sqrt{\ell(\ell+1)}\boldsymbol{Y}_{\ell m}^{E}(\theta,\phi)
\end{equation}
are the eigenfunctions of the uncoupled sphere modes, Eq.(\ref{Psinlm}), 
repeated here for convenience. After the integration over the angular part
this integral reduces, in the case of Schenberg antenna, to

\begin{equation}
    f^{GW}_{n2m}=\frac 12\ddot h_m(t)M_SR\sqrt{\frac{3}{5\pi}}\int_0^1 \xi^3(A_{n2}(\xi R)+3B_{n2}(\xi R))d\xi
    =\frac 12\ddot h_m(t)M_S\chi_n R,
    \label{fn2m}
\end{equation}

where

\begin{equation}
    \chi_n=\sqrt{\frac{3}{5\pi}}\int_0^1 \xi^3(A_{n2}(\xi R)+3B_{n2}(\xi R))d\xi
    \label{chi}.
\end{equation}

For the Schenberg antenna we have $\chi_1=-0.6004$.

\section{The Detector Model}
\label{sec:ITM}

As discussed above, the mechanical oscillations of the Schenberg antenna are monitored by a set of parametric transducers coupled on its surface. From a mathematical point of view, Johnson and Merkowitz \cite{Merkowitz97} proposed a model in which the output data from six transducers coupled to the antenna surface are related by decomposing them into the quadrupolar modes of the sphere. This method allows the reconstruction of the parameters that characterize the incident GW. 

The movement equation for the displacement vector field $\bm u(\bm x,t)$ of 
a solid subjected to external forces density $\bm f(\bm x,t)$ is given 
by \cite{Landau07}
\begin{equation}
    \rho\frac{\partial^2\bm u}{\partial t^2}-
    (\lambda_L+\mu_L)\bm\nabla(\bm{\nabla\cdot u})-
    \mu_L\bm\nabla^2\bm u=\bm f,
    \label{eqmov}
\end{equation}

where $\lambda_L$ and $\mu_L$ are the tangential and volumetric Lamé coefficients of the material respectively. The initial conditions are
$\bm u(\bm x,0)=0$ and $\dot{\bm u}(\bm x,0)=0$.
The solution of (\ref{eqmov}) is obtained expanding the displacement vector $\bm u(\bm x,t)$
in series of the eigenfunctions $\bm\Psi_N(\bm x)$ of the equation

\begin{equation}
    (\lambda_L+\mu_L)\bm\nabla(\bm{\nabla\cdot\Psi}(\bm x))+
    \mu_L\bm\nabla^2\bm\Psi(\bm x) = -\rho^2\bm\Psi(\bm x)
    \label{eigenoperator}
\end{equation}

subjected to the boundary condition of tension free at the surface of the sphere \cite{Maggiore}

\begin{equation}
    \lambda_L(\bm{\nabla\cdot u})\bm{\hat r}+
    2\mu_L(\bm{\hat r\cdot\nabla})\bm u+
    \mu_L\bm{\hat r\times}(\bm{\nabla\times u})=0.
\end{equation}

The displacement vector field can be expanded as
\begin{equation}
    \bm u(\bm x,t)=\sum_N a_N(t)\bm\Psi_N(\bm x),
    \label{uxt}
\end{equation}

where $N$ is a set of indices, $a_N(t)$ is the time-dependent mode amplitude and $\bm \Psi_N$ obeys the normalization condition

\begin{equation}
    \int_V\rho(\bm x)\bm\Psi_N(\bm x)\cdot\bm\Psi_{N'}(\bm x)d^3x=M_S\delta_{NN'}.
    \label{normcond}
\end{equation}

The integration is over the volume $V$ of the sphere.
After substituting (\ref{eigenoperator}) and (\ref{uxt}) in (\ref{eqmov}),
multiplying by $\bm\Psi_{N'}$ and integrating over the volume of the sphere 
using (\ref{normcond}), we obtain

\begin{equation}
    M_S\ddot a_N(t)+\kappa_Sa_N(t)=\int_V\bm\Psi_N(\bm x)\cdot\bm f(\bm x,t)d^3x
    \label{eqmova}
\end{equation}

with $\kappa_S$ being the elastic constant. 

At this point it is convenient to introduce a damping term in Eq.(\ref{eqmova})

\begin{equation}
    M_S\ddot a_N(t)+C_S\dot a_N(t)+\kappa_Sa_N(t)=
    \int_V\bm\Psi_N(\bm x)\cdot\bm f(\bm x,t)d^3x,
    \label{eqmovad}
\end{equation}

where $C_S=w_N/Q_N$, $w_N$ the natural angular frequency of mode $N$ and $Q_N$ the mechanical quality factor $Q$ for mode $N$. 
The values of the parameters are given in Tab.(\ref{parameters}).

\subsection{The uncoupled sphere}

The solution of (\ref{eigenoperator}) subjected to the boundary condition of 
tension free at its surface are the natural modes of the sphere.
They consist of two families of solution, the toroidal modes $\bm\Psi_{n\ell m}^T$ and the spheroidal modes $\bm\Psi_{n\ell m}$ (see \cite{Lobo95}). We rewrite here this solution in terms of the vector spherical harmonics defined in Sec.(\ref{vsharmonics}). Regarding the toroidal modes, in the case of a coupled sphere, they do not impart radial motion on the transducers, and the Schenberg detector is not sensitive to them, besides the fact that GWs do not excite these modes.

\subsubsection{Spheroidal modes}\label{spheroidal}

The spheroidal modes are given by

\begin{equation}
    \boldsymbol{\Psi}_{n\ell m}(\boldsymbol{x})=
    A_{n\ell}(r)\boldsymbol{Y}_{\ell m}^{L}(\theta,\phi)+
    B_{n\ell}(r)\sqrt{\ell(\ell+1)}\boldsymbol{Y}_{\ell m}^{E}(\theta,\phi),
    \label{Psinlm}
\end{equation}

where

\begin{align}
    A_{n\ell}(r)&=C_{n\ell}\left[\beta_{3}(k_{n\ell}R)j'_{\ell}(q_{n\ell}r)-
    \ell(\ell+1)\frac{q_{n\ell}}{k_{n\ell}}\beta_{1}(q_{n\ell}R)
    \frac{j_{\ell}(k_{n\ell}r)}{k_{n\ell}r}
    \right]\\
    B_{n\ell}(r)&=C_{n\ell}\left[\beta_{3}(k_{n\ell}R)\frac{j_{\ell}(q_{n\ell}r)}{q_{n\ell}r}-
    \frac{q_{n\ell}}{k_{n\ell}}\beta_{1}(q_{n\ell}R)\beta_5(k_{n\ell}r)\right].
\end{align}

The transverse wave vectors $k_{n\ell}$, the longitudinal wave vectors
$q_{n\ell}$ and the natural angular frequencies $w_{n\ell}=2\pi f_{n\ell}$ are
the solution of the system of equations

\begin{align}
    {\rm det}\left[\begin{array}{cc}
        \beta_{4}(qR) & \ell(\ell+1)\beta_{1}(kR) \\
        \beta_{1}(qR) & \beta_{3}(kR)
    \end{array}\right] &=0 \label{det}\\
    qc_l &=w\label{q}\\
    kc_t &= w\label{k},
\end{align}

where betas are given by

\begin{align}
    \beta_{0}(z) &= \frac{j_{\ell}(z)}{z^{2}}\label{beta0}\\
    \beta_{1}(z) &= \frac{d}{dz}\left(\frac{j_{\ell}(z)}{z}\right)\label{beta1}\\
    \beta_{2}(z) &= \frac{d^2j_{\ell}(z)}{dz^2}\label{beta2}\\
    \beta_{3}(z) &= \frac{1}{2}\beta_{2}(z)+
        \left(\frac{\ell(\ell+1)}{2}-1\right)\beta_{0}(z)\label{beta3}\\
    \beta_{4}(z) &= \beta_{2}(z)-\frac{\sigma}{1-2\sigma}j_{\ell}(z)\label{beta4}\\
    \beta_{5}(z) &= \frac 1z\frac{d}{dz}(zj_\ell(z)).
\end{align}

The coefficients $c_l$ and $c_t$ are respectively the 
longitudinal

\begin{equation}
    c_l=\sqrt{\frac{\mu_L}{\rho}}\sqrt{\frac{2-2\sigma}{1-2\sigma}}
    \label{cl}
\end{equation}
and transversal 
\begin{equation}
    c_t = \sqrt{\frac{\mu_L}{\rho}}
    \label{c_t}
\end{equation}

velocities of the elastic waves. We define the ratio

\begin{equation}
    \delta = \frac{c_l}{c_t}.
    \label{r}
\end{equation}

Here, $\rho$ is the density of the sphere and
$\sigma$ the Poisson ratio.
The Poisson ratio can be written in terms of the ratio of the longitudinal and transversal sound velocities

\begin{equation}
    \sigma = \frac 12\frac{\delta^2-2}{\delta^2-1}.
    \label{sigmafr}
\end{equation}

The solution of the system of equations 
(\ref{det}, \ref{q}, \ref{k}) only depends on $c_l$ and $c_t$, in this way using the measured values of the monopole and quadrupole frequencies we were able to determine them. The results are given in Tab.(\ref{parameters}). 

The relationship between the Poisson ratio and the Young modulus $E$ 
with the Lamé coefficients $\lambda_L$ and $\mu_L$ are

\begin{equation}
    \frac{\lambda_L}{\mu_L}=\frac{2\sigma}{1-2\sigma}\qquad \mu_L=\frac{E}{2(1+\sigma)}.
    \label{lambdasmu}
\end{equation}

\subsection{Antenna parameters at 4 K}

The linear thermal expansion as a function of temperature is given by \cite{Ashcroft,Reif}
\begin{equation}
    \alpha_{\rm lin}(T)=\alpha_0\frac{\rho}{3BA}\left(\gamma c_V^{\rm ion}(T)+\frac 23 c_V^{\rm el}(T) \right),
    \label{alpha}
\end{equation}

where $\alpha_0$ is a
constant such that $\alpha_{\rm lin}(273.15)=1.75\times 10^{-5}\rm K^{-1}$ \cite{RobertHandbook},
$A$ is the weighted average of CuAl6 atomic mass in kg, $B$ is the bulk modulus

\begin{equation}
    B = \frac{E}{3(1-2\sigma)}=\frac{2\rho c_t^2(1+\sigma)}{3(1-2\sigma)}
    \label{B}
\end{equation}

and $\gamma$ is the weighted average of the CuAl6 Gruneisen coefficient. 
The lattice specific heat is

\begin{equation}
    c_V^{\rm ion}(T)=3R_Gf_D\left(\frac{\Theta_D}{T}\right),
    \label{cVion}
\end{equation}

where $R_G$ is the gas constant, $\Theta_D$ is the weight average of CuAl6 Debye's temperature. The Debye's function is

\begin{equation}
    f_D(y)=\frac{3}{y^3}\int_0^y\frac{{\rm e}^x x^4}{({\rm e}^x-1)^2}dx.
    \label{fD}
\end{equation}

The electrons specific heat is given by

\begin{equation}
    c_V^{\rm el}(T) = R_G\frac{\pi^2}{2}\frac{T}{T_F} 
    \label{cVel}
\end{equation}
with $T_F$ being the weight average of CuAl6 Fermi temperature. Then 
the radius at 4 K will be given by

\begin{equation}
    R=R_0+R_0\int_{300}^4\alpha_{lin}(T)dT.
    \label{R}
\end{equation}

After calculating $c_l$ and $c_t$, based on its measured values at 300 K and 2 K and using the frequency of the monopolar mode and the mean frequency of the quadrupolar modes,
we are able to calculate the radius of the sphere at 4 K. The solution must take into account that the coefficient of linear expansion depends on the Poisson's ratio as well as the equations (\ref{det}, \ref{q}, \ref{k}) depends on it.
With this methodology it is possible to calculate physical constants of CuAl6.
The results are given in Tab.(\ref{parameters}).

\begin{table}
\centering
\caption{Parameters of the Schenberg antenna.} 
\vspace{2pc}
\resizebox{15cm}{!}{
\begin{tabular}{l|l|l} \hline 
{\bf Description} & {\bf Value} & {\bf Method} \\\hline\hline
Quadrupole frequencies at 2 K & 3172.485, 3183.000, 3213.623, 3222.900, 3240.000 $\pm$ $0.001$Hz & 
measured\\\hline
Quadrupole frequencies at 300 K& 3045, 3056, 3086, 3095, 3102 $\pm$ $0.5\,\rm Hz$ & measured\\\hline
Monopole frequency at 300 K & $f_{10}=6443.0\pm 0.5\,\rm Hz$ & measured\\ \hline
Antenna's radius at 300 K & $R_0=0.3233\,\rm m$ & measured\\\hline
Antenna mass & $M_S=1124\rm\, kg$ & measured  \\ \hline 
Antenna's density at 300 K & $\rho=7938.523\pm 19\,\rm kg/m^3$ & measured\\\hline
Transducer first stage mass &  $M_1=59.7100\pm 0.5\,\rm mg$ & measured \\ \hline
Transducer second stage mass &  $M_2=12.0\pm 0.5\, \rm mg$ & measured  \\ \hline\hline 
Monopole frequency at 4 K & $f_{10}=6713.42\,\rm Hz$ & calculated\\ \hline
Mean quadrupole frequency at 4 K & $\bar f_{12}=3205.94\,\rm Hz$ & calculated\\\hline
Longitudinal sound velocity at 4 K &$c_l=4937.6\,\rm m/s $& calculated using (\ref{det}, \ref{q}, \ref{k})\\\hline
Transversal sound velocity at 4 K & $c_t=2448.2\,\rm m/s$ & calculated using (\ref{det}, \ref{q}, \ref{k})\\\hline 
Linear thermal expansion coefficient at 273.15 K & $\alpha_0=1.75\times 10^{-5}\,\rm K^{-1}$ & reference  \cite{RobertHandbook}\\\hline
Weight average of CuAl6 Debye temperature & $\Theta_D=319.74\,\rm K$ & reference \cite{Ashcroft}\\\hline
Weight average of CuAl6 Fermi temperature &$T_F=84449.46\,\rm K$ & reference \cite{Ashcroft}\\\hline
Weight average CuAl6 Gruneisen coefficient &$\gamma=1.912$ & reference \cite{Callen}\\\hline
Sound velocities ratio & $r=2.016847$ & calculated using (\ref{r}) \\\hline
Poisson ratio & $\sigma=0.337010$ & calculated using (\ref{sigmafr}) \\\hline
Sphere radius as 4 K & $R = 0.32213\,\rm m$ & calculated using (\ref{det}, \ref{q}, \ref{k}, \ref{R}) \\\hline  
Sphere density at 4 K & $\rho=8025.04\,\rm kg/m^3$ & calculated \\\hline
Volumetric Lamé coefficient & $\mu_L=48.100\,\rm GPa$ & calculated using (\ref{c_t}) \\\hline
Tangential Lamé coefficient & $\lambda_L=99.455\,\rm GPa$ & calculated using (\ref{lambdasmu}) \\\hline
Young modulus & $E = 128.621\,\rm GPa$ & calculated using (\ref{lambdasmu}) \\\hline
Bulk modulus & $B=131.522\,\rm GPa$ & calculated using (\ref{B}) \\\hline
Chi factor &  $\chi=-0.6004$ & calculated using (\ref{chi})\\   \hline
Radial component factor at $r=R$ & $\alpha=2.88345$ & $\alpha=A_{12}(R)$\\\hline 
Antenna equivalent mass & $M_{\rm eq}=340\,\rm kg$ & $M_{\rm eq}=
\frac{4\pi}{5\alpha^2}M_S$\\\hline
Antenna effective mass & $M_{\rm eff}=283\,\rm kg$ & 
$M_{\rm eff}=\frac{5}{6}M_{\rm eq}$\\ \hline
Transducer amplification factor & ${\rm amp}=\sqrt{\frac{M_{\rm eff}}{M_2}}=4740$\\\hline
\end{tabular}} 
\label{parameters}
\end{table}

\subsection{The antenna coupled with transducers}

In order to detect GWs, six two stage transducers are coupled to the Schenberg antenna \cite{Paula15}.
Each stage of the transducers has the same resonance frequency of the first quadrupole mode $f_0=3205.94\rm Hz$ and are sensitive only to the radial movement of the sphere.
Transducers are devices that monitor the motion of the antenna surface. 
If a hypothetical GW excites the sphere quadrupolar modes, 
the corresponding mechanical energy will be transferred from the antenna to the transducers. 
Jonhson \& Merkowitz \cite{Johnson93} discovered that if we use six transducers and locate each of them at the center of a pentagonal face of a truncated icosahedron projected onto one hemisphere of the sphere, then by a suitable linear combination 
of the output of the transducers, the so called mode channels, 
we can obtain a direct correspondence between the spherical
amplitudes $h_m(t)$ of the GW and the quadrupolar modes of the sphere $a_{2m}(t)$. The angles of each of these transducers are given in Tab.(\ref{angles}).

\begin{table}\label{TabAngles}
\centering
\caption{Polar and azimuthal angles  $(\theta,\phi)$ of the transducers positions, $\varphi=(1+\sqrt{5})/2$.}
    \label{angles} 
\vspace{2pc}
\begin{tabular}{c|c|c} \hline
{\bf Transducer} & $\theta$ & $\phi$ \\ \hline\hline  
T3 & ${\rm acos}\left(\frac{1}{\sqrt{3}\varphi\sqrt{\varphi+2}}\right)=79.18^{\rm o}$ & $0^{\rm o}$ \\ \hline
T6 & ${\rm acos}\left(\frac{\varphi+1}{\sqrt{3}\sqrt{\varphi+2}}\right)=
37.37^{\rm o}$ & $60^{\rm o}$ \\ \hline
T2 & ${\rm acos}\left(\frac{1}{\sqrt{3}\varphi\sqrt{\varphi+2}}\right)=79.18^{\rm o}$ & $120^{\rm o}$ \\ \hline
T5 & ${\rm acos}\left(\frac{\varphi+1}{\sqrt{3}\sqrt{\varphi+2}}\right)=
37.37^{\rm o}$ & $18^{\rm o}$ \\ \hline
T1 & ${\rm acos}\left(\frac{1}{\sqrt{3}\varphi\sqrt{\varphi+2}}\right)=79.18^{\rm o}$ & $24^{\rm o}$ \\ \hline
T4 & ${\rm acos}\left(\frac{\varphi+1}{\sqrt{3}\sqrt{\varphi+2}}\right)=
37.377^{\rm o}$ & $30^{\rm o}$ \\ \hline
\end{tabular}
\end{table}

The Schenberg antenna makes use of two-modes parametric transducers. 
In this model the transducer motion is exclusively radial and only the $m$ 
quadrupole modes are of interest. 
In an homogeneous sphere the modes are degenerated but 
in the real antenna they are not. 

\begin{figure}[!ht]
\begin{center}
   \includegraphics[scale=0.3]{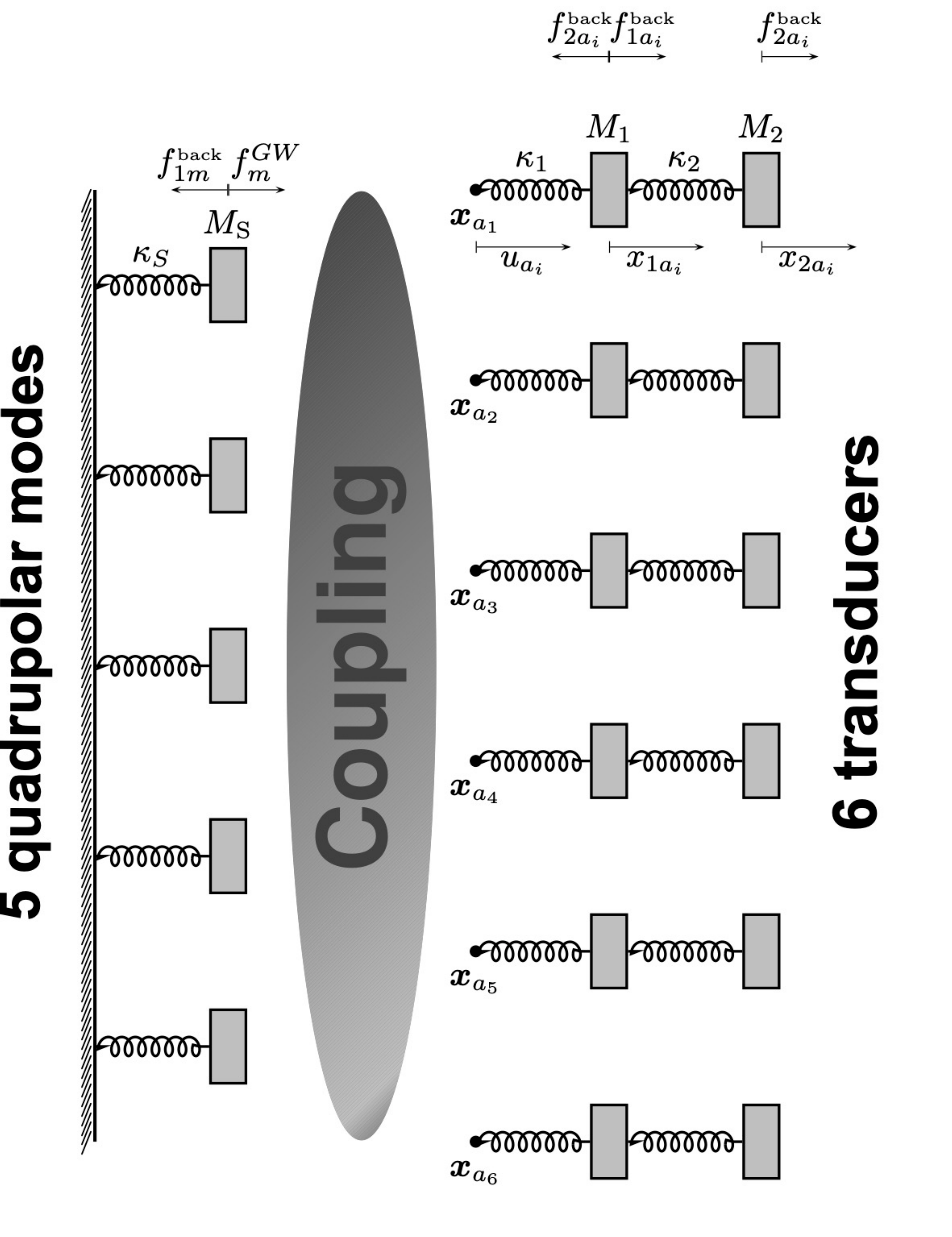}
     \caption{Schematic drawing representing in 2D the 3D coupling of the first five quadrupole (independent) modes of the Schenberg spherical antenna (left) with the six two-mode transducers (right). Each transducer more or less couples with each quadrupole mode of the sphere, depending on its position on the surface of the sphere in relation to the quadrupole mode in question. Due to these couplings, each transducer outputs information from all 17 modes.}
     \label{molas}
     \end{center}
\end{figure}

The forces acting on the sphere (Fig. \ref{molas}) are the GW force given by

\begin{equation}
    \bm f^{GW}=\sqrt{\frac{4\pi}{15}}\rho r\ddot h_m(t)
    \left(\bm Y_{2m}^L+\frac{\sqrt{6}}{2}\bm Y_{2m}^E\right),
\end{equation}

the spring back reaction of the six transducers over the sphere at the positions $\bm x_{a}$

\begin{equation}
	\bm f_1^\kappa=\sum_{a=1}^6\kappa_1(x_{1a}-u_a)\delta(\bm x-\bm x_a){\bm e}_a,
\end{equation}

the damping back reaction of the resonators of the six transducers over the sphere at the positions $\bm x_{a}$

\begin{equation}
	\bm f_1^C=\sum_{a=1}^6C_1(\dot x_{1a}-\dot u_a)\delta(\bm x-\bm x_a){\bm e}_a,
\end{equation}

where $C_{1}$ is the damping term of the first resonator. The noise back reaction forces from the resonators are

\begin{equation}
	\bm f_1^{back}=\sum_{a=1}^6f_{1a}^{back}\delta(\bm x-\bm x_a){\bm e}_a,
\end{equation}

where $x_{1a}$ is the displacement of the first resonator from its equilibrium position, 
${\bm e}_a$ is the radial unit vector at the position $\bm x_a$ over the sphere and 
$u_a$ the deformation of the sphere at $\bm x_a$ given by (repeated here for convenience)

\begin{equation}
	u_a = \sum_{m=-2}^2a_m(t)\bm\Psi_m(\bm x_a)\cdot{\bm e}_a.
\end{equation}

The equation for $\bm\Psi_m$, Eq.(\ref{Psinlm}), is rewritten here with $n=1$, $\ell=2$,
$A_{12}(r)=\alpha(r)$ and $B_{12}=\beta(r)$
\begin{equation}
    \boldsymbol{\Psi}_{m}(\boldsymbol{x})=
    \alpha(r)\boldsymbol{Y}_{m}^{L}(\theta,\phi)+
    \beta(r)\sqrt{6}\boldsymbol{Y}_{m}^{E}(\theta,\phi),
    \label{Psim}
\end{equation}

so that we have for $u_a$

\begin{equation}
	u_a=\alpha(R)\sum_{m=-2}^2a_m(t)Y_m(\theta_a,\phi_a)=
	\alpha(R)\sum_{m=-2}^2a_m(t)B_{ma}.
\end{equation}

In matrix notation this is

\begin{equation}
	{\bf u}=\alpha{\bf B}^T{\bf a},
\end{equation}

where $\alpha=\alpha(R)$ and the bold letters are matrices in which each entry represents a transducer.
The movement equation for the displacement of the sphere
surface $\bf u$ is given in Sec.(\ref{moveqsd}).

The forces over the first resonator are the noise forces 
between the first resonator and the sphere, $f_1^{back}$, 
the back action of the noise force between resonator 1 and 2, $-f_2^{back}$,
the spring 2 back action over the resonator 1, $f_2^\kappa=\kappa_2(x_{2a}-x_{1a})$,
the damping back action of spring 2, $f_2^C=C_2(\dot x_{2a}-\dot x_{1a})$,
the reaction of the spring 1, $-f_1^\kappa=-\kappa_1(x_{1a}-u_a)$,
the damping of the spring 1, $-f_1^C=-C_1(\dot x_{1a}-\dot u_a)$,
The forces over the resonator 2 are the noise force between resonator 1 and 2, $f_2^{back}$,
the reaction of the spring 2, $-f_2^\kappa=-\kappa_1(x_{2a}-x_{1a})$,
the damping of the spring 2, $-f_2^C$.
The equations for the system are

\begin{align}
	M_S\ddot a_m(t) &= - C_S\dot a_m(t) -\kappa_S a(t) + 
	\int\bm\Psi_m(\bm x)\cdot \bm f(\bm x,t)d^3x\label{eqa}\\
	M_1\ddot x_{1a} &= f_{1a}^{back}-f_{2a}^{back} - \kappa_1(x_{1a}-u_a) 
		- C_1(\dot x_{1a}-\dot u_a)
		+ \kappa_2(x_{2a}-x_{1a})
		+ C_2(\dot x_{2a}-\dot x_{1a})\\
	M_2\ddot x_{2a} &= f_{2a}^{back} -\kappa_2(x_{2a}-x_{1a})-
	C_2(\dot x_{2a}-\dot x_{1a}),
\end{align}

where $\bm f(\bm x,t)=\bm f_1^\kappa(\bm x,t)+\bm f_1^C(\bm x,t)-\bm f_1^{back}(\bm x,t)+\bm f^{GW}(\bm x,t)$ 
are the surface forces over the sphere and the GW force.
The transducers frequencies are tuned with the frequency of the quadrupole mode of the homogeneous sphere $w_0$ such that

\begin{equation}
	\frac{\kappa_S}{M_S}=\frac{\kappa_1}{M_1}=\frac{\kappa_2}{M_2}=w_0^2.
\end{equation}

For the real antenna we take $w_0$ as the mean value of the measured 
quadrupole mode frequencies $w_m$.
For the maximum energy transfer from the sphere to the resonators the masses obeys the relation \cite{Richard84}

\begin{equation}
	\frac{M_1}{M_{\rm eff}}=\frac{M_2}{M_1}=\mu^2,
\end{equation}

where the effective mass of the antenna $M_{\rm eff}$ is calculated in the Appendix (\ref{SecMeff}).
The integral in Eq.(\ref{eqa}) can be written as

\begin{align}
	\int\bm\Psi_m(\bm x)\cdot \bm f(\bm x,t)d^3x =&
	\int\bm\Psi_m\cdot\bm f_1^\kappa d^3x+
	\int\bm\Psi_m\cdot\bm f_1^Cd^3x-\nonumber\\
	&\int\bm\Psi_m\cdot\bm f_1^{back}d^3x+
	\int\bm\Psi_m\cdot\bm f_m^{GW} d^3x.
\end{align}

The first integral on the right hand side gives

\begin{equation}
	\int\bm\Psi_m\cdot\bm f_1^\kappa d^3x=\kappa_1\alpha\sum_{a=1}^N Y_m(\theta_a,\phi_a)q_{1a}=
	\kappa_1\alpha\sum_{a=1}^N B_{ma}q_{1a}=\kappa_1\alpha[{\bf B}{\bf q}_1]_m.
\end{equation}

Similarly the second gives

\begin{equation}
	\int\bm\Psi_m\cdot\bm f_1^C d^3x=
	C_1\alpha\sum_{a=1}^N Y_m(\theta_a,\phi_a)\dot q_{1a}=
	C_1\alpha[{\bf B}{\bf\dot q}_1]_m
\end{equation}

and the third

\begin{equation}
	\int\bm\Psi_m\cdot\bm f_1^{back} d^3x=
	\alpha\sum_{a=1}^N Y_m(\theta_a,\phi_a)f_{1a}^{back}=
	\alpha[{\bf B}{\bf f}_1^{back}]_m,
\end{equation}

where $q_{1a}=x_{1a}-u_a$ and $q_{2a}=x_{2a}-x_{1a}$,
the fourth is the Eq.(\ref{chi}). The result is

\begin{equation}
	\int\bm\Psi_m(\bm x)\cdot \bm f(\bm x,t)d^3x=
	\kappa_1\alpha[{\bf B}{\bf q}_1]_m+
	C_1\alpha[{\bf B}{\bf\dot q}_1]_m-
	\alpha[{\bf B}{\bf f}_1^{back}]_m+
	{f}_m^{GW}(t).
\end{equation}

From now on we will use the column matrix

\begin{equation}
	{\bf f}^{GW}(t)=\begin{pmatrix} f_{-2}^{GW}(t) \\ f_{-1}^{GW}(t) \\f_0^{GW}(t) \\f_1^{GW}(t) \\f_2^{GW}(t) \end{pmatrix}.
\end{equation}

The equations in the new variables and in matrix notation are

\begin{align}
    M_S{\bf\ddot a} &+ C_S{\bf\dot a}+\kappa_S{\bf a}-
		C_1\alpha{\bf B}{\bf\dot q}_1
		-\kappa_1\alpha{\bf B}{\bf q}_1 = {\bf f}^{GW}-
		\alpha{\bf B}{\bf f}_1^{back}\nonumber\\
	M_1\alpha{\bf B}^T{\bf\ddot a} &+M_1{\bf\ddot q}_1+
    M_1{\bf\ddot q}_1+C_1{\bf\dot q_{1}}-C_2{\bf\dot q_{2}} +
    \kappa_1{\bf q}_{1}-\kappa_2{\bf q}_{2} = {\bf f}_{1}^{back}-
    {\bf f}_2^{back}\\
    M_2\alpha{\bf B}^T{\bf\ddot a} &+
    M_2{\bf\ddot q}_2+M_2{\bf\ddot q}_1+
    M_2{\bf\ddot q}_1+M_2{\bf\ddot q_{2}}+
    C_2{\bf\dot q_{2}}-C_1{\bf\dot q_{1}} +
    \kappa_2{\bf q}_{2}-\kappa_1{\bf q}_{1} = 
    {\bf f}_{2}^{back}\nonumber.
\end{align}

In block matrix notation we have

\begin{align}
   \left[
        \begin{array}{ccc}
            M_S\,{\bf I}              & {\bf 0}         & {\bf 0} \\
            M_1\alpha {\bf B}^T   & M_1{\bf I}  & {\bf 0} \\
            M_2\alpha {\bf B}^T   & M_2{\bf I}  & M_2{\bf I}
        \end{array}\right]
        \left[
        \begin{array}{c}
            {\bf \ddot a}\\ {\bf\ddot q}_1 \\ {\bf \ddot q}_2 
        \end{array}\right] &+
        \left[
        \begin{array}{ccc}
            {\rm Diag}({C_i}_S)      & -C_1\alpha{\bf B} & {\bf 0} \\
            {\bf 0}   & C_1{\bf I}        & -C_2{\bf I} \\
            {\bf 0}   & {\bf 0}               & C_2{\bf I}
        \end{array}\right]
        \left[
        \begin{array}{c}
            {\bf \dot a}\\ {\bf\dot q}_1 \\ {\bf \dot q}_2 
        \end{array}\right] \nonumber\\ &+
        \left[
        \begin{array}{ccc}
            {\rm Diag}({k_i}_S)  & -k_1\alpha{\bf B} & {\bf 0} \\
            {\bf 0}   & k_1{\bf I}        & -k_2{\bf I} \\
            {\bf 0}   & {\bf 0}               & k_2{\bf I}
        \end{array}\right]
        \left[
        \begin{array}{c}
            {\bf a}\\ {\bf q}_1 \\ {\bf q}_2 
        \end{array}\right]=
        \left[
        \begin{array}{ccc}
            {\bf I}  & -\alpha{\bf B} & {\bf 0} \\
            {\bf 0}   & {\bf I}       & -{\bf I} \\
            {\bf 0}   & {\bf 0}       & {\bf I}
        \end{array}\right]
        \left[
        \begin{array}{c}
            {\bf f}^{GW}\\ {\bf f}_{1}^{back} \\ {\bf f}_{2}^{back} 
        \end{array}\right].
\end{align}

These equations can be rewritten in terms of the block 
matrices

\begin{equation}
 {\sfb M}'{\sfb{\ddot q}}+{\sfb C}'{\sfb{\dot q}}+
 {\sfb K}'{\sfb q}={\sfb P}{\sfb f}.
 \label{EqMov1}
\end{equation}

From  now on we use sanserif boldface letters for 
block matrices.
Here $\sfb q$ is the displacement matrix
\begin{equation}
    {\sfb q}=\left[
        \begin{array}{c}
            {\bf a}\\ {\bf q}_1 \\ {\bf q}_2 
        \end{array}\right],
\end{equation}

where ${\bf a}_{5\times 1}$ is the antenna's mode amplitude,
${\bf q}_{1\, 6\times 1}$ and ${\bf q}_{2\,6\times 1}$ 
are vectors of the relative displacements for 
resonator 1 and resonator 2 of each transducer. 
The mass matrix is

\begin{equation}
    {\sfb M}'=\left[
        \begin{array}{ccc}
            M_S\,{\bf I}              & {\bf 0}         & {\bf 0} \\
            M_1\alpha {\bf B}^T   & M_1{\bf I}  & {\bf 0} \\
            M_2\alpha {\bf B}^T   & M_2{\bf I}  & M_2{\bf I}
        \end{array}\right],
\end{equation}

where ${\bf B}_{5\times 6}$ is the model matrix.
Let us rewrite this matrix in term of the effective mass using the ratios $\mu$ and

\begin{equation}
    \nu^2=\frac{M_S}{M_{\rm eff}}.
\end{equation} 

We have

\begin{equation}
    {\sfb M}'=M_{\rm eff}\left[
        \begin{array}{ccc}
            \nu^2\,{\bf I}              & {\bf 0}         & {\bf 0} \\
            \mu^2\alpha {\bf B}^T     & \mu^2{\bf I}  & {\bf 0} \\
            \mu^4\alpha {\bf B}^T     & \mu^4{\bf I}  & \mu^4{\bf I}
        \end{array}\right]=M_{\rm eff}{\sfb M}
\end{equation}

\begin{equation}
    {\sfb K}'=\left[
        \begin{array}{ccc}
            {\rm Diag}({k_i}_S)  & -k_1\alpha{\bf B} & {\bf 0} \\
            {\bf 0}   & k_1{\bf I}        & -k_2{\bf I} \\
            {\bf 0}   & {\bf 0}               & k_2{\bf I}
        \end{array}\right]
\end{equation}
or
\begin{equation}
    {\sfb K}'=M_{\rm eff}w_0^2\left[
       \begin{array}{ccc}
       \nu^2{\rm Diag}\left(\frac{w_m^2}{w_0^2}\right)  & -\mu^2\alpha{\bf B} & {\bf 0} \\
            {\bf 0}   & \mu^2{\bf I}        & -\mu^4{\bf I} \\
            {\bf 0}   & {\bf 0}               & \mu^4{\bf I}
       \end{array}\right]=M_{\rm eff}w_0^2{\sfb K}
\end{equation}
\begin{equation}
    {\sfb C}'=\left[
        \begin{array}{ccc}
            {\rm Diag}({C_i}_S)      & -C_1\alpha{\bf B} & {\bf 0} \\
            {\bf 0}   & C_1{\bf I}        & -C_2{\bf I} \\
            {\bf 0}   & {\bf 0}               & C_2{\bf I}
        \end{array}\right].
        \label{ClFDT}
\end{equation}

As we will see in Sec.(\ref{Seq.PSD}), it will be convenient to write this matrix as

\begin{equation}
    {\sfb C}'=M_{\rm eff}\left[
        \begin{array}{ccc}
            {2\nu^2\rm Diag}({\beta_m}) & -2\mu^2\beta_1\alpha{\bf B} & {\bf 0} \\
            {\bf 0}   & 2\mu^2\beta_1{\bf I}        & -2\mu^4\beta_2{\bf I} \\
            {\bf 0}   & {\bf 0}   & 2\mu^4\beta_2{\bf I}
        \end{array}\right],
\end{equation}

where $2\beta_m=w_m/Q$, $2\beta_1=w_0/Q_1$ and $2\beta_2=w_0/Q_2$, with $Q_1$ and $Q_2$ the quality factors of the resonators (Fig. \ref{molas}). This matrix can yet be written as

\begin{equation}
    {\sfb C}'=M_{\rm eff}\frac{w_0}{Q}\left[
        \begin{array}{ccc}
        \nu^2{\rm Diag}\left(\frac{w_mQ}{w_0Q_m}\right) & -\mu^2\frac{Q}{Q_1}\alpha{\bf B} & {\bf 0} \\
            {\bf 0}   & \mu^2\frac{Q}{Q_1}{\bf I}        & -\mu^4\frac{Q}{Q_2}{\bf I} \\
            {\bf 0}   & {\bf 0}               & \mu^4\frac{Q}{Q_2}{\bf I}
        \end{array}\right]=M_{\rm eff}\frac{w_0}{Q}{\sfb C}.
\end{equation}

At this point we know that

\begin{equation}
	{\rm Diag}\left(\frac{w_n}{w_0}\right)\approx
	{\rm Diag}\left(\frac{w_n^2}{w_0^2}\right)\approx{\bf I},
	\label{approxw}
\end{equation}

and if we approximate

\begin{equation}
    \frac{Q}{Q_m}=1 \qquad\frac{Q}{Q_1}=1 \qquad\frac{Q}{Q_2}=1
    \label{approxQ}
\end{equation}

we get

\begin{equation}
    {\sfb C}=\left[
        \begin{array}{ccc}
        \nu^2{\bf I} & -\mu^2 \alpha{\bf B} & {\bf 0} \\
            {\bf 0}   & \mu^2 {\bf I}        & -\mu^4 {\bf I} \\
            {\bf 0}   & {\bf 0}               & \mu^4 {\bf I}
        \end{array}\right].
\end{equation}

Then is justified to put ${\sfb C}={\sfb K}$. 
The matrix $\sfb P$ is

\begin{equation}
    {\sfb P}=\left[
        \begin{array}{ccc}
            {\bf I}  & -\alpha{\bf B} & {\bf 0} \\
            {\bf 0}   & {\bf I}       & -{\bf I} \\
            {\bf 0}   & {\bf 0}       & {\bf I}
        \end{array}\right]
\end{equation}

and its inverse

\begin{equation}
    {\sfb P}^{-1}=\begin{bmatrix}
        {\bf I} & \alpha{\bf B} & \alpha{\bf B}\\
        {\bf 0} & {\bf I} & {\bf I}\\
        {\bf 0} & {\bf 0} & {\bf I}
    \end{bmatrix}.
\end{equation}

The force matrix is

\begin{equation}
    {\sfb f}=\left[
        \begin{array}{c}
            {\bf f}^{GW}\\ {\bf f}_{1}^{back} \\ {\bf f}_{2}^{back} 
        \end{array}\right].
\end{equation}

The movement equation then reads

\begin{equation}
 M_{\rm eff}{\sfb M}{\sfb{\ddot q}}+
 M_{\rm eff}\frac{w_0}{Q}{\sfb K}{\sfb{\dot q}}+
 M_{\rm eff}w_0^2{\sfb K}{\sfb q}={\sfb P}{\sfb f}.
\end{equation}

We will need to diagonalize the matrix 
${\sfb M}^{-1}{\sfb K}$, but this matrix is not symmetric. 
In order to symmetrize it we change the coordinates defining 
${\sfb q}={\sfb N}{\sfb y}$ where 
\begin{equation}
    {\sfb N}=\left[
        \begin{array}{ccc}
            {\bf I}/\nu  & {\bf 0}        & {\bf 0} \\
            {\bf 0}   & {\bf I}/\mu   & {\bf 0} \\
            {\bf 0}   & {\bf 0}                & {\bf I}/\mu^2
        \end{array}\right]
\end{equation}

and pre-multiply by $\sfb N$

\begin{equation}
 M_{\rm eff}\sfb{NMN\ddot y}+
 M_{\rm eff}\frac{w_0}{Q}\sfb{NKN\dot y}+
 M_{\rm eff}w_0^2\sfb{NKNy}=\sfb{NPf}.
\end{equation}

Multiplying both sides of the equation by $\sfb{(NMN)}^{-1}$ and defining $2\beta=\frac{w_0}{Q}$ we get

\begin{equation}
 M_{\rm eff}\sfb{\ddot y}+
 2\beta M_{\rm eff}\sfb{(NMN)}^{-1}\sfb{NCN\dot y}+
 M_{\rm eff}w_0^2\sfb{(NMN)}^{-1}\sfb{NKNy}=
 \sfb{(NMN)}^{-1}\sfb{NPf}.
\end{equation}

Let us define the variables ${\sfb M}_y$, ${\sfb K}_y$ and 
${\sfb P}_y$, where the subscript is the indicative that these matrices are of the equation for $\sfb y$. The equation then reads

\begin{equation}
 M_{\rm eff}\sfb{\ddot y}+
 2\beta M_{\rm eff}\sfb{C}_y\sfb{\dot y}+
 M_{\rm eff}w_0^2\sfb{K}_y\sfb{y}={\sfb P}_y\sfb{f},
\end{equation}

where

\begin{equation}
{\sfb C}_y=\sfb{(NMN)}^{-1}\sfb{NCN} =
\begin{pmatrix}
{\rm diag}\frac{w_nQ}{w_0Q_n} & -\gamma{\bf B} & {\bf 0}\\
-\gamma{\bf B}^T{\rm diag}\frac{w_nQ}{w_0Q_n} & 
    \left(\frac{3\gamma^2}{2\pi}+1\right){\bf I}-\frac{\gamma^2}{4\pi}{\bf 1}  & -\mu{\bf I}\\
{\bf 0} & -\mu{\bf I} & (\mu^2+1){\bf I}
\end{pmatrix}
\end{equation}
\begin{equation}
{\sfb K}_y=\sfb{(NMN)}^{-1}\sfb{NKN} =
\begin{pmatrix}
{\rm diag}\frac{w_n^2}{w_0^2} & -\gamma{\bf B} & {\bf 0}\\
-\gamma{\bf B}^T{\rm diag}\frac{w_n^2}{w_0^2} & 
    \left(\frac{3\gamma^2}{2\pi}+1\right){\bf I}-\frac{\gamma^2}{4\pi}{\bf 1} & -\mu{\bf I}\\
{\bf 0} & -\mu{\bf I} & (\mu^2+1){\bf I}
\end{pmatrix},
\end{equation}

where $\gamma=\alpha\mu/\nu$ and ${\bf 1}$ is a matrix full of ones

\begin{equation}
    \sfb{(NMN)}^{-1} =
        \begin{pmatrix}
            {\bf I} & {\bf 0} & {\bf 0}\\
            -\gamma{\bf B}^T & {\bf I} & {\bf 0}\\
            {\bf 0} & -\mu & {\bf I}
        \end{pmatrix}
\end{equation}
\begin{equation}
    {\sfb P}_y=\sfb{(NMN)}^{-1}\sfb{NP} =
    \begin{pmatrix}
        \frac 1\nu{\bf I} & -\frac\gamma\mu {\bf B} & {\bf 0}\\
        -\frac{\gamma}{\nu}{\bf B}^T & 
        \left(\frac{3\gamma^2}{2\pi\mu}+\frac 1\mu\right){\bf I}-\frac{\gamma^2}{4\pi\mu}{\bf 1} & -\frac 1\mu{\bf I}\\
        {\bf 0} & -{\bf I} & \left(1+\frac 1{\mu^2}\right){\bf I}
    \end{pmatrix}.
\end{equation}

At this point it is necessary to do some approximations. We can see that ${\sfb K}_y$ is not symmetric, but we also know that ${\rm diag}(w_n^2/w_0^2)\approx {\bf I}$. So in the entry 
${\sfb K}_{y21}$ we approximate 
${\rm diag}(w_n^2/w_0^2)={\bf I}$. On the other side, 
if we want to diagonalize the damping matrix 
${\sfb C}_y$ with the same matrix 
${\sfb U}$ that diagonalize ${\sfb K}_y$ we do the approximations $Q/Q_n=1$ and 
${\rm diag}(w_n/w_0)={\bf I}$ in the entry ${\sfb C}_{y21}$ 
and $Q/Q_n=w_n/w_0$ in the entry ${\sfb C}_{y11}$
then ${\sfb C}_y={\sfb K}_y$ and both matrices are diagonalized with the same matrix ${\sfb U}$. The equation then reads

\begin{equation}
  M_{\rm eff}\sfb{\ddot y}+
  M_{\rm eff}2\beta{\sfb K}_y\sfb{\dot y}+
  M_{\rm eff}w_0^2\sfb K_y\sfb y=
  {\sfb P}_y\sfb f.
\end{equation}

To diagonalize ${\sfb K}_y$ using the modal matrix $\sfb U$, we define $\sfb y={\sfb U}\sfb z$, 
pre multiply both sides of the equation by ${\sfb U}^T$ and take the Fourier transform. The
result is

\begin{equation}
  -M_{\rm eff}w^2\sfb{\tilde z}+
  M_{\rm eff}2\beta jw\sfb{D\tilde z}+
  M_{\rm eff}w_0^2\sfb{D\tilde z}=
  {\sfb U}^T{\sfb P}_y\sfb{\tilde f},
\end{equation}

where $\sfb D$ is the diagonal matrix
given by ${\sfb D}={\sfb U}^T{\sfb K}_y{\sfb U}$ and the tilde letters are the
Fourier transform of its corresponding variables. We omit the $w$ dependence 
in some cases to leave the notation cleaner.
If we define the diagonal matrix

\begin{equation}
    {\sfb L}(w)=
         (-M_{\rm eff}w^2{\sfb I}+
         M_{\rm eff}2\beta jw{\sfb D}+
         M_{\rm eff}w_0^2{\sfb D})
\end{equation}

we get

\begin{equation}
    {\sfb L}(w)\sfb{\tilde z}(w)=
    {\sfb U}^T{\sfb P}_y\sfb{\tilde f}(w)
    \label{LZ}.
\end{equation}

We invert to find $\boldsymbol{\tilde z}$

\begin{equation}
  \sfb{\tilde z}={\sfb L}^{-1}(w){\sfb U}^T
  {\sfb P}_y\sfb{\tilde f},
\end{equation}

where

\begin{equation}
    {\sfb L}^{-1}(w)=\frac{1}{M_{\rm eff}}
    {\rm Diag}\left(\frac{1}{-w^2+(2j\beta w+w_0^2)D_{11}},
    \cdots,\frac{1}{-w^2+(2j\beta w+w_0^2)D_{1717}} \right).
\end{equation}

Returning to the old variables we have

\begin{equation}
    \sfb{\tilde q}=
    \sfb{NU}{\bf L}^{-1}(w){\sfb U}^T{\sfb N}^{-1}{\sfb M}^{-1}
    {\sfb P}\sfb{\tilde f}.
    \label{qtilde}
\end{equation}

The transfer functions for the input $\sfb{\tilde f}$ will be

\begin{equation}
    {\sfb G}(w)=\sfb{NU}\sfb{L}^{-1}(w){\sfb U}^T
    {\sfb N}^{-1}{\sfb M}^{-1}{\sfb P},
\end{equation}

where the block matrix $\sfb G$ can be written as

\begin{equation}
    {\sfb G}=\begin{pmatrix}
        {\bf G}_{00} & {\bf G}_{01} &{\bf G}_{02} \\
        {\bf G}_{10} & {\bf G}_{11} &{\bf G}_{12} \\
        {\bf G}_{20} & {\bf G}_{21} &{\bf G}_{22}
    \end{pmatrix}.
\end{equation}

Then, we can write Eq.(\ref{qtilde}) as

\begin{equation}
    \begin{pmatrix} 
        {\bf \tilde a} \\ {\bf\tilde q}_1 \\ {\bf\tilde q}_2
    \end{pmatrix}=
    \begin{pmatrix}
        {\bf G}_{00} & {\bf G}_{01} &{\bf G}_{02} \\
        {\bf G}_{10} & {\bf G}_{11} &{\bf G}_{12} \\
        {\bf G}_{20} & {\bf G}_{21} &{\bf G}_{22}
    \end{pmatrix}
    \begin{pmatrix}
        {\bf\tilde f}_0 \\{\bf\tilde f}_1\\{\bf\tilde f}_2
    \end{pmatrix}.
\end{equation}

\subsection{Classical noise power spectrum matrix}\label{Seq.PSD}
In this work we will assume that the noise is an ergodic wide sense 
stationary stochastic process being analysed in an interval of time $T_o$.
Let $x(t)$ with Fourier transform $\tilde x(w)$ be a process satisfying these conditions, then the Power Spectral Density (PSD) of $x$ is calculated as 
(see Whalen Chap.(2) \cite{Whalen} and Maggiore \cite{Maggiore} for details)

\begin{equation}
    S_{xx}=E[{\tilde x}(w)\tilde x(w)^*]T_o.
\end{equation}

Our system is contaminated with forces of thermal noise $\bm f_{\rm th}$,
forces of back action on the membrane 
${\bm f}_{\rm bk}$,
series forces 
$\bm f_{\rm se}$ and phase forces 
$\bm f_{\rm ph}$.
The measured quantity is the output ${\bf q}_2$ (transducer membrane) of our system

\begin{equation}
	{\bf\tilde q}_2={\bf G}_{20}{\bf\tilde f}_0+
	{\bf G}_{21}{\bf\tilde f}_1+
	{\bf G}_{22}{\bf\tilde f}_2+
	{\bf G}_{22}{\bf\tilde f}_{bk}+
	{\bf\tilde f}_{se}+{\bf\tilde f}_{ph}.
\end{equation}

The PSD of the output ${\bf q}_2$ is, assuming that the noise forces
of different kind are non correlated and
the forces $\bf\tilde f$ are of thermal origin

\begin{align}
	{\sfb S}_{qq} &=
	{\bf G}_{20}E[{\bf\tilde f}_0
	{\bf\tilde f}_0^\dagger]{\bf G}_{20}^\dagger+
	{\bf G}_{21}E[{\bf\tilde f}_1
	{\bf\tilde f}_1^\dagger]{\bf G}_{21}^\dagger+
	{\bf G}_{22}E[{\bf\tilde f}_2
	{\bf\tilde f}_2^\dagger]{\bf G}_{22}^\dagger+
	{\bf G}_{22}E[{\bf\tilde f}_{bk}
	{\bf\tilde f}_{bk}^\dagger]{\bf G}_{22}^\dagger+
	E[{\bf\tilde f}_{se}{\bf\tilde f}_{se}^\dagger]+
	E[{\bf\tilde f}_{ph}{\bf\tilde f}_{ph}^\dagger]\nonumber\\
	&=
	{\bf G}_{20}{\bf S}_{f_0f_0}{\bf G}_{20}^\dagger+
	{\bf G}_{21}{\bf S}_{f_1f_1}{\bf G}_{21}^\dagger+
	{\bf G}_{22}{\bf S}_{f_2f_2}{\bf G}_{22}^\dagger+
	{\bf G}_{22}{\bf S}_{bk}{\bf G}_{22}^\dagger+
	{\bf S}_{se}+{\bf S}_{ph}.
\end{align}

The thermal noise power spectrum is based on the 
fluctuation dissipation theorem that stays that given a system with
equation
\begin{equation}
	{\sfb L}(w)\sfb{\tilde z}=\sfb{\tilde f}
\end{equation}
the power spectrum of the fluctuation force $\sfb f$ is given by

\begin{equation}
	{\sfb S}_{ff}=4k_BT{\rm Re}[\bm{{\cal Z}}(w)],
\end{equation}

where $\bm{{\cal Z}}$ is the impedance of the system given by

\begin{equation}
	\bm{{\cal Z}}(w) = \frac{{\sfb L}(w)}{jw}.
\end{equation}

In our case we have

\begin{equation}
	{\sfb S}_{th}=4k_BT{\rm Re}
	\left[\frac{{\sfb L}(w)}{jw}\right].
\end{equation}

But from Eq.(\ref{EqMov1}) and Eq.(\ref{ClFDT}) we have

\begin{equation}
    {\sfb S}_{th}=
	4k_BT{\sfb P}^{-1}{\sfb C}'=4k_BT
    \begin{pmatrix}
        M_S\frac{w_0}{Q}{\bf I}_{5\times 5} & {\bf 0} & {\bf 0}\\
        {\bf 0} & M_1\frac{w_0}{Q_1}{\bf I}_{6\times 6} & {\bf 0}\\
        {\bf 0} & {\bf 0} & M_2\frac{w_0}{Q_1}{\bf I}_{6\times 6}
    \end{pmatrix}=
    \begin{pmatrix}
        {\bf S}_{f_0f_0} & {\bf 0} & {\bf 0}\\
        {\bf 0} & {\bf S}_{f_1f_1} & {\bf 0}\\
        {\bf 0} & {\bf 0} & {\bf S}_{f_2f_2}
    \end{pmatrix}
    \qquad [{\rm N^2/Hz}].
\end{equation}

The back action noise force acting on the membrane is \cite{Tobar00}

\begin{equation}
	{\bf S}_{bk}=\frac{P^2_{\rm inc}S_{\rm am}}{2w_p^2}
	\left(\frac{2Q_e}{f_p}\frac{df}{dx}\right)^2
	{\bf I}_{6\times 6}
	\qquad [{\rm N^2/Hz}],
\end{equation}

the series noise acting directly on the output is

\begin{equation}
	{\bf S}_{se}=\frac{(T_{\rm amp}+T)k_B}{P_{\rm inc}}
	\left(\frac{2Q_e}{f_p}\frac{df}{dx}\right)^{-2}
	{\bf I}_{6\times 6}
	\qquad [{\rm m^2/Hz}]
\end{equation}

and the phase noise also acting directly on the output is

\begin{equation}
	{\bf S}_{ph}=S_{pph}
	\left(\frac{2\pi}{w}\frac{df}{dx}\right)^{-2}
	{\bf I}_{6\times 6}
    \qquad [{\rm m^2/Hz}].
\end{equation}

\subsection{Standard Quantum Limit Noise}

In the following section we will derive the expression of the standard quantum noise. This will allow us to obtain the standard quantum limit of the Schenberg detector. The power signal-to-noise ratio $\rho^2$
for an optimum filter (matched filter) is \cite{Wainshtein62}

\begin{equation}
    \rho^2=\frac{1}{2\pi}\int_{-\infty}^{\infty}\frac{|M(w)|^2}{S_{nn}^{\rm ds}(w)}dw,
\end{equation}

where $M(w)$ is the Fourier transform of the signal of interest and $S_{nn}^{\rm ds}(w)$
the double side power spectral density of the noise. Our signal is the vector with the spherical amplitudes ${\bf h}(t)$.
Using the single side power spectral density matrix ${\bf S}_{nn}(w)$, the
expression of the power signal-to-noise ratio becomes

\begin{equation}
    \rho^2=\frac{4}{2\pi}\int_0^{\infty}
    {\bf\tilde h}^\dagger(w){\sfb S}_{nn}^{-1}(w)
    {\bf\tilde h}(w)dw.
\end{equation}

For bursts of duration $\tau_g\approx 1\,\rm ms$ the maximum 
bandwidth frequency is $\Delta f_{\rm max}\approx 1\,\rm kHz$ and 
${\bf\tilde h}(w)$ does not change very much
from its value at the resonant frequency $f_0$ in the band 
$\Delta f$ of the detector.
We can define a mean power spectral density $\bar S_{nn}$ such that this integral can be approximated by

\begin{equation}
  \rho^2=
  \frac{4\Delta w}{2\pi}|{\bf\tilde h}^\dagger(w_0)|
    \left(\frac{1}{\Delta w}\int_0^{\infty}
        {\bf\hat{\tilde h}}^\dagger
        {\sfb S}_{nn}^{-1}(w)
        {\bf\hat{\tilde h}}dw
    \right)|{\bf\tilde h}(w_0)|
    =
  \frac{4|{\bf\tilde h}(w_0)|^2\Delta f}{\bar S_{nn}},
  \label{rho2}
\end{equation}

where ${\bf h}(w_0)=|{\bf h}(w_0)|{\bf\hat{\tilde h}}$,

\begin{equation}
    |{\bf h}(w_0)|=\sqrt{\sum_{m=-2}^2h_m^2(w_0)}
\end{equation}
and
\begin{equation}
    \bar S_{nn}=\left(\frac{1}{\Delta w}\int_0^{\infty}
        {\bf\hat{\tilde h}}^\dagger
        {\sfb S}_{nn}^{-1}(w)
        {\bf\hat{\tilde h}}dw
    \right)^{-1}.
\end{equation}

We can obtain $|{\bf h}(w_0)|$ as a function of the energy deposited by the burst 
on the sphere using the formula \cite{Landau01}
\begin{equation}
    E_s=\frac{1}{2M}\left|\int_{-\infty}^{\infty}f(t){\rm e}^{jw_0 t}dt\right|^2,
\end{equation}

where $f(t)$ is the external force acting on the harmonic oscillator and $M$ its mass.
Starting from the movement equation for the sphere modes (Eq.\ref{eqmova}), the mass of the mode is $M_S$ as a result of the normalization condition and the force is 
$f(t)=\frac 12M_S\chi R\ddot h_m(t)$.
The integration gives for each mode $m$

\begin{equation}
    E_{sm}=\frac 18M_S\chi^2R^2w_0^4|\tilde h_m(w_0)|^2,
\end{equation}

while the energy deposited in all modes is

\begin{equation}
    E_s=\frac 18M_S\chi^2R^2w_0^4|\bm{\tilde h}(w_0)|^2.
    \label{Es}
\end{equation}

The sensitivity is obtained when $\rho^2=1$. With this value, comparing Eq.(\ref{rho2}) with Eq.(\ref{Es})

we obtain the mean power density spectrum as a function of the energy deposited in the sphere

\begin{equation}
    \bar S_{nn}=\frac{32E_s}{M_S\chi^2 R^2w_0^4}.
\end{equation}

The energy deposited in terms of the number of phonons $n$ is $E_s=n\hbar w_0$.
The sensitivity at the quantum limit is when $n=1$

\begin{equation}
    S_{\rm SQL}=\frac{32\hbar\Delta f}{M_S\chi^2 R^2w_0^3}.
\end{equation}

We can write this expression as a function of the longitudinal sound velocity
and the longitudinal wave vector for the quadrupolar mode 
$w_0=w_{12}=q_{12}v_{\shortparallel}$

\begin{equation}
    S_{\rm SQL}=\frac{32\hbar\Delta f}{M_S\chi^2(q_{12}R)^2v_{\shortparallel}^2w_0}.
\end{equation}

For Schenberg at $4\,\rm K$, $f_0=f_{12}=3205.94\,\rm Hz$, $M_S=1124\,\rm kg$, $\chi=-0.6004$, 
$R=32.214\,\rm cm$ and $\Delta f=110\,\rm Hz$.
With these values the spectral amplitude is

\begin{equation}
    h_S(w_0)=\sqrt{S_{\rm SQL}}=3.29\times 10^{-23}\,\sqrt{\rm Hz^{-1}}.
\end{equation}

\subsection{Sensitivity for Classical Noise}
The spectral amplitude $h_S(w)$ represents the input
GW spectrum that would produce a signal equal to the noise spectrum
observed at the output of the antenna instrumentation.

A useful way to characterize the sensitivity of a GW detector is to calculate
the $h_S(w)$ such that with optimal filtering the signal to 
noise ratio

\begin{equation}
	\rho^2=\frac{1}{2\pi}\int_{-\infty}^\infty \sigma(w)dw
\end{equation}

is equal to 1 for each bandwidth. Here

\begin{equation}
	\sigma(w)=\bm{\tilde q}_2^\dagger
	{\bf S}^{-1}_{qq}\bm{\tilde q}_2,
\end{equation}

where $\bm q_2$ are the output of the second transducer's resonators,
$\dagger$ stands for Hermitean conjugate. 
The sensitivity of the detector is obtained by searching for an input GW with amplitude ${\bf h}$ that mimics the thermal noise at the output, with $\rho = 1$ per bandwidth. In other words we search for an 
${\bf h}$ such that

\begin{equation}
    \bm{\tilde q}_{2}^{\dagger}{\bf S}_{qq}^{-1}
    {\bm{\tilde q}_2}=1
\end{equation}

or

\begin{equation}
    X(w)^2
    {\bf \tilde h}^\dagger{\bf T_v}^T
    {\bf G}_{20}^\dagger(w){\bf S}_{qq}^{-1}
    {\bf G}_{20}(w){\bf T_v}{\bf\tilde h}=1.
\end{equation}

As we do not know the polarization neither the direction of the incoming wave we take the mean over all angles

\begin{equation}
    \int_{\psi=0}^\pi \int_{\phi=0}^{2\pi}\int_{\theta=0}^\pi
    {\bf \tilde h}^\dagger{\bf T_v}^T
    {\bf G}_{20}^\dagger(w){\bf S}_{qq}^{-1}
    {\bf G}_{20}(w){\bf T_v}{\bf\tilde h}
    \frac{\sin\theta d\psi d\theta d\phi}{4\pi^2}=
    \frac{1}{X(w)^2}.
\end{equation}

Then we obtain

\begin{equation}
    \frac 15({\tilde h}_+^2+{\tilde h}_\times^2)
    {\rm Tr}({\bf G}_{20}^\dagger(w){\bf S}_{qq}^{-1}
    {\bf G}_{20}(w)) 
    =\frac{1}{X(w)^2}
\end{equation}

and the amplitude spectral density $h_S(w)=\sqrt{{\tilde h}_+^2+{\tilde h}_\times^2}$

\begin{equation}
    h_S(w)=\frac{\sqrt{5}}
    {\sqrt{X(w)^2
    {\rm Tr}({\bf G}_{20}^\dagger(w){\bf S}_{qq}^{-1}
    {\bf G}_{20}(w))}}.
\end{equation}

\begin{figure}[!ht]
\begin{center}
   \includegraphics[scale=1]{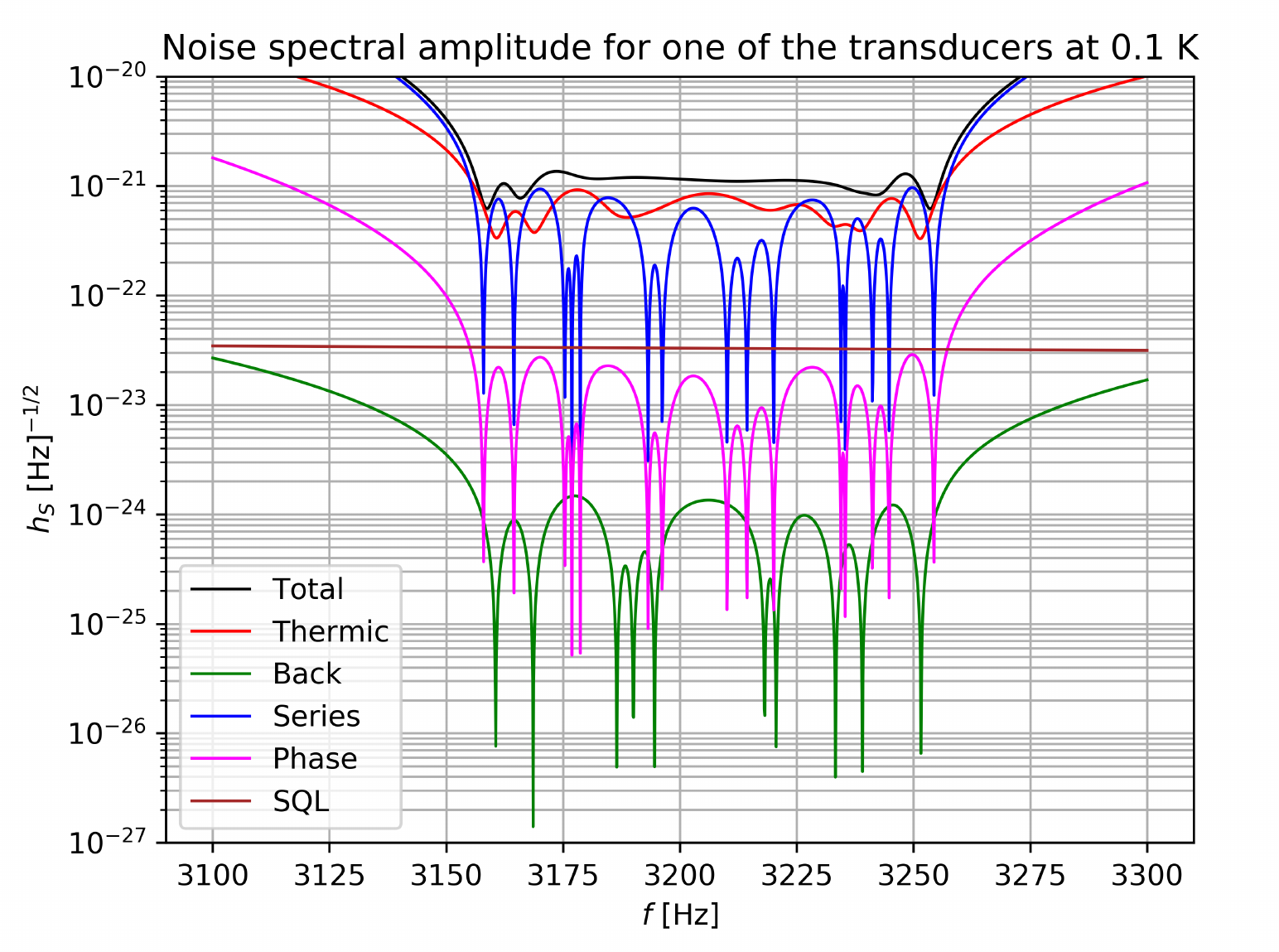}
     \caption{Sensitivity curves of the various type of noises for one of the six transducers of the Schenberg antenna at T = 0.1 K.}
     \label{FighS}
     \end{center}
\end{figure}

\begin{figure}[!ht]
\begin{center}
   \includegraphics[scale=1]{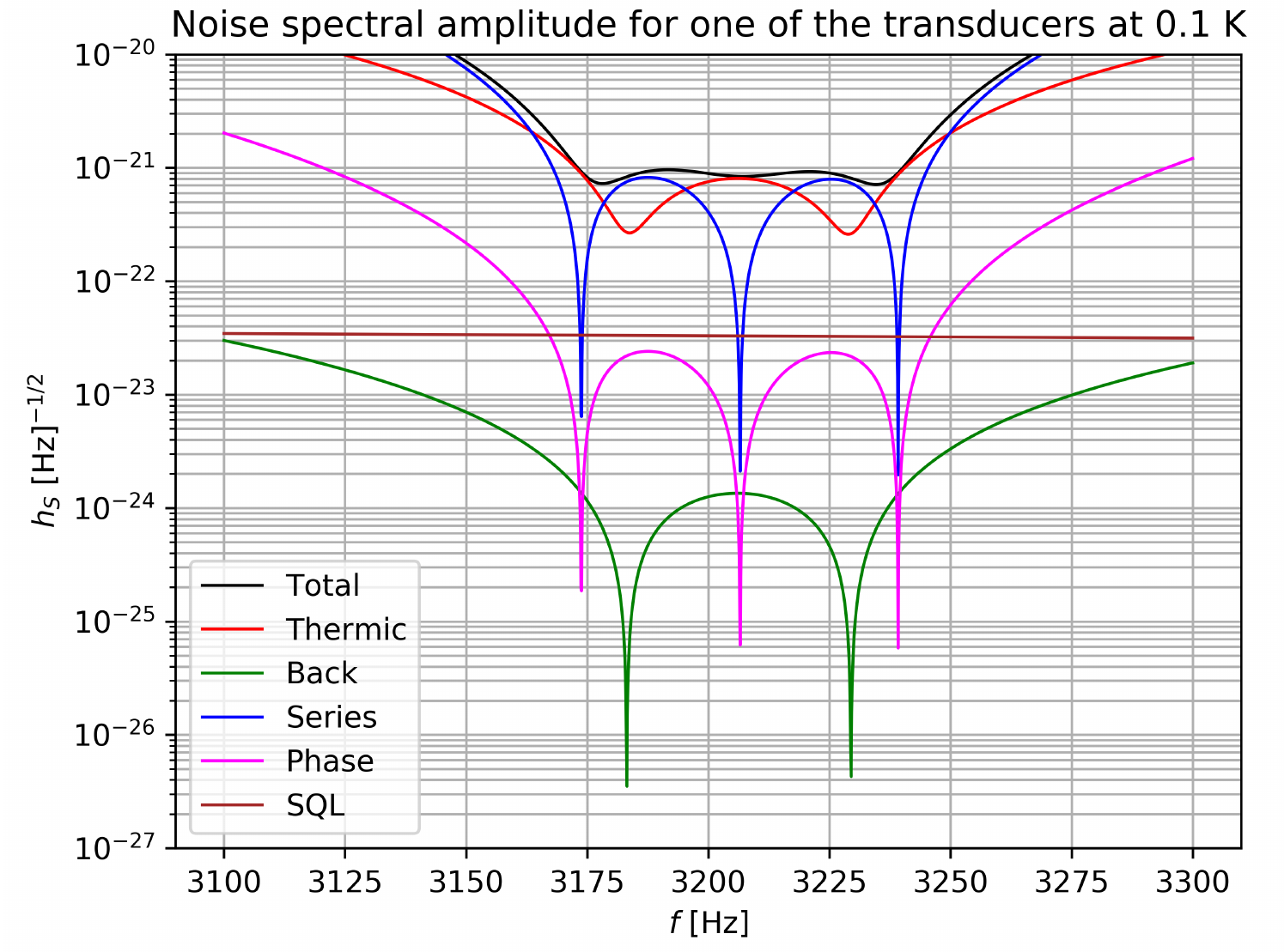}
     \caption{Sensitivity curves of the various type of noises for one of the six transducers of the Schenberg antenna system at T = 0.1 K for the sphere degenerated case.}
     \label{FighSd}
     \end{center}
\end{figure}

The sensitivity curves for various kind of noises 
for each of the six transducers of the real antenna are shown in Fig.(\ref{FighS}) 
using the parameters given in Tab.(\ref{FigParam}). In the case of a degenerated sphere the sensitivity curves for each of the six transducers would be as in Fig.(\ref{FighSd}).

\begin{table}
\centering
\caption{Parameters used in the sensitivity curve.} 
\vspace{2pc}
\begin{tabular}{l|l} \hline 
{\bf Description} & {\bf Value} \\\hline\hline
Temperature  & $T=100$\,mK\\\hline 
Sphere mechanical Q & $Q=1\times 10^7$ \\\hline
Resonator 1 mechanical Q & $Q_1=1\times 10^6$ \\\hline
Resonator 2 mechanical Q & $Q_2=1\times 10^5$ \\\hline
Transducer central frequency & $F_T=3206.3$\,Hz\\\hline
Transducer minus frequency & $F_-=3172.5$\,Hz\\\hline
Transducer plus frequency & $F_+=3240.0$\,Hz\\\hline
Pump frequency & $F_{\rm pump}=1\times 10^{10}$\,Hz\\\hline
Electric coupling constant & $\beta_{\rm e}=0.65$\\\hline
Frequency shift with distance & $\frac{df}{dx}=7.26\time 10^{14}\,\rm Hz/m$\\\hline
Oscillator incident power & $P_{\rm inc}=1\times 10^{-10}$\,W\\\hline
Noise amplifier temperature & $T_{\rm amp}=10$\,K\\\hline
Electrical quality factor & $Q_{\rm e}=3.8\times 10^5$\\\hline
Phase noise spectral density & $S_{\rm p}=1\times 10^{-13}$\,dBc/$\rm Hz$\\\hline
Amplitude noise spectral density & 
$S_{\rm a}=1\times 10^{-14}$\,dBc/$\rm Hz$\\\hline
Amplifier loss & $L_{\rm amp}=5$\\\hline\hline
\end{tabular}
\label{FigParam}
\end{table} 

The sensitivity of the Schenberg antenna will be better than the sensitivity of each transducer. Assuming that all transducers have the same sesitivity, the sensitivity of the Schenberg antenna ($h_S$) would be $(1/h_S)^2 = (1/h_{T1})^2 + (1/h_{T2})^2 + (1/h_{T3})^ 2 + (1/h_{T4})^2 + (1/h_{T5})^2 + (1/h_{T6})^2 = 6 \times(1/h_T)^2$, which implies that $h_S = h_T/ \sqrt{6}$.

\section{Discussions and Conclusions}
\label{sec:disc}

The calculation of the Schenberg antenna design sensitivity for each of the sphere six transducers was revised in this work taking into account both the degenerate (perfect sphere) and the non-degenerate sphere (quadrupole modes with their different frequencies), due to the symmetry break caused by the machining of the holes for the fixation of the transducers and the copper rod for the sphere suspension. As usual, all noises are referenced at the “input of the sphere” where the oscillating movement of the sphere surface occurs.

The dominant noises are the Brownian and the series noise, taking into account the parameters available for this initial version of the Schenberg antenna. For an advanced version of the Schenberg antenna (aSchenberg), which would reach the standard quantum limit of it ($3.29\times 10^{- 23}  sqrt{Hz}^{- 1}$), the sensitivity at each of the six transducers would be $\sqrt{6}$ times this or ($\sim 8\times 10^{- 23}  sqrt{Hz}^{- 1}$). To achieve this sensitivity at each niobium transducer we have to replace them with sapphire or silicon transducers, and with niobium coating in the microwave cavity region. In this way, we could reach mechanical quality factors of the order of $10^{8}$ \cite{Locke}. The sphere would have to undergo annealing or be replaced by another material, such as beryllium copper. Values of mechanical Qs close to $10^{8}$ have already been reached by Frossati (1996) \cite{Frossati} for small copper-beryllium spheres.

Series noise can be minimized by rounding the edges of the transducer microwave klystron cavities, using a niobium deposition with less than 100 parts per million impurities, to increase the already achieved 380k electrical quality factor by a factor of 10 or more. The loss $L_{\rm amp}$ in the microwave transmission line  that carry the signal from the transducer to the cryogenic amplifier (the first line of amplifiers in the system) would need to be reduced by a factor of 5. This could be achieved using niobium coaxial cables. Finally, the electronics used in the cryogenic amplifiers would need to be replaced by one that would reduce the noise temperature from 10 K to 1 K, at the operating frequency of 10 GHz.

All these modifications, necessary to reach the standard quantum limit, are challenging, but not impossible to achieve for the small spherical antenna of 0.65 cm in diameter. As parametric transducers are used, it would be possible to perform signal squeezing and exceeds the standard quantum limit, but this would require higher mechanical and electrical Qs and even less noisy electronics, which starts to be unfeasible or doubtful to be achieved.

Note, however, that the sensitivity achieved by aLIGO in the O3 run has already reached the standard quantum limit of this spherical antenna, therefore, the only reasonable justification for remounting the Schenberg antenna and trying to place it in the sensitivity of the standard quantum limit would be to detect gravitational waves with another physical principle, different from the one used by laser interferometers. This other physical principle would be the absorption of the gravitational wave energy by a resonant mass. The question that arises, then, is whether gravitational wave signals reach Earth with sufficient amplitude to be detected by the spherical antenna operating at the standard quantum limit. To answer this question, we are analyzing aLIGO's O3 data in the range where the Schenberg antenna is most sensitive: 3.15 kHz to 3.26 kHz, looking for any type of signal (burst, chirp, continuous or stochastic). We  look  forward  to  providing  the results of this investigation in the near future.

\appendix\section{Effective mass}
\label{SecMeff}
The effective mass of the antenna that a transducer sees according to Zhou \cite{Zhou1995} 
is the mass that placed on the surface of the sphere acquires the same energy that the sphere has. 
Let us start calculating the equivalent mass for the quadrupole modes.
The kinetic energy of the antenna for the quadrupole mode $m$ is

\begin{equation}
    E_k=\frac 12\int\rho\Psi_{m}^2\dot a_{m}^2d^3x=\frac 12 M_S\dot a_{m}^2.
    \label{Ekeq1}
\end{equation}

The velocity of the surface at the position of the transducer $a$ in radial direction for the mode $m$ is

\begin{equation}
    v_{rma}=\dot a_m\bm{\Psi}_m\cdot\bm{e}_a=
    \dot a_m\alpha Y_m(\theta_a,\phi_a).
    \label{vra}
\end{equation}

The kinetic energy for the effective mass of the mode $m$ in this position is

\begin{equation}
    E_k=\frac 12M_m\alpha^2Y_m^2(\theta_a,\phi_a) \dot a_m^2.
    \label{Ekeq2}
\end{equation}

Comparing (\ref{Ekeq1}) and (\ref{Ekeq2}) we have the equation for the effective mass for the mode $m$

\begin{equation}
	M_S=M_m\alpha^2Y_m^2(\theta_a,\phi_a) 
\end{equation}

and rearranging the equation we have

\begin{equation}
	\frac{M_S}{\alpha^2M_m}=Y_m^2(\theta_a,\phi_a). 
\end{equation}

The sum over all modes gives

\begin{equation}
\frac{M_{S}}{\alpha^{2}}\sum_{m=-2}^{2}\frac{1}{M_m}=
\sum_{m=-2}^{2}Y_{m}^{2}(\theta_a,\phi_a).
\end{equation}

The sum rule for spherical harmonics for the quadrupole gives

\begin{equation}
\sum_{m=-2}^{2}Y_{m}^{2}(\theta,\phi)=\frac{5}{4\pi}.
\end{equation}

If we define the equivalent mass for the modes of the sphere
as

\begin{equation}
\frac{1}{M_{\rm eq}}=\frac{1}{M_1}+\frac{1}{M_2}+\frac{1}{M_3}+\frac{1}{M_4}+\frac{1}{M_5}
\end{equation}

then

\begin{equation}
M_{{\rm eq}}=\frac{4\pi}{5\alpha^{2}}M_{S}
\end{equation}

and using $M_{S}=1124$ kg and $\alpha^{2}=8.28584$ we have



\begin{equation}
M_{{\rm eq}}=0.30228M_{S}=340.934\,{\rm kg}.
\end{equation}

As we show explicitly in the next section the effective mass for $N$ transducers considering five modes is given by

\begin{equation}
    M_{\rm eff} =\frac 5NM_{\rm eq}\quad\text{for six transducers}\quad M_{\rm eff} =284.111\,{\rm kg}.
    \label{Meff}
\end{equation}

\subsection{Explicit effective mass calculation for {\it N} transducers}

\subsection{For {\it N=1}}

With the transducer at the position $a$ we have

\begin{equation}
    \frac 12{M_S}\sum_{m=-2}^2\dot a_m^2=
    \frac 12M_{\rm eff_1}v_a^2
    \label{Ek1}.
\end{equation}

The velocity of the sphere surface at this position is

\begin{equation}
    v_a=\alpha\sum_{m=1}^5\dot a_m Y_m(\theta_a,\phi_a).
\end{equation}

The square is given by

\begin{equation}
    v_a^2=\alpha^2\sum_{m,n=1}^5\dot a_m\dot a_n Y_m(\theta_a,\phi_a)Y_n(\theta_a,\phi_a)
\end{equation}

and the mean over the angles is

\begin{equation}
	\overline{v_a^2}=\alpha^{2}\sum_{m,n=1}^{5}\dot{a}_{m}\dot{a}_{n}\frac{1}{4\pi}
	\underbrace{\int Y_{m}(\theta,\phi)Y_{n}(\theta,\phi)\sin\theta d\theta}_{\delta_{mn}}=
	\frac{\alpha^2}{4\pi}\sum_{m=-2}^{2}\dot{a}_{m}^{2}.
	\label{meanva2}
\end{equation}

The mean of both sides of Eq.(\ref{Ek1}) results
\begin{equation}
	\frac 12{M_S}\sum_{m=-2}^{2}\dot a_m^2=
	\frac 12M_{\rm eff_1}\overline{v_a^2}=
	\frac 12M_{\rm eff_1}\frac{\alpha^2}{4\pi}\sum_{m=-2}^{2}\dot{a}_{m}^{2},
\end{equation}

such that

\begin{equation}
	M_{\rm eff_1}=\frac{4\pi}{\alpha^2}M_S=
	\frac{5}{1}\frac{4\pi}{5\alpha^2}M_S=\frac{5}{1}M_{\rm eq}=1705\,\rm kg.
\end{equation}

\subsection{For {\it N} transducers}
If we have $N$ transducers the kinetic energy is function of 
\begin{equation}
	\bar v_1^2+\bar v_2^2+\cdots+\bar v_N^2=
	N\frac{\alpha^2}{4\pi}\sum_{m=-2}^{2}\dot{a}_{m}^{2}
\end{equation}

such that

\begin{equation}
	M_{\rm eff_N}=\frac 1N\frac{4\pi}{\alpha^2}M_S=
	\frac{5}{N}\frac{4\pi}{5\alpha^2}M_S=\frac{5}{N}M_{\rm eq}.
\end{equation}

\subsection{For six transducers in truncated icosahedron configuration}
In matrix notation Eq.(\ref{Ek1}) can be written

\begin{equation}
	\frac 12M_S{\bf\dot a}^T{\bf\dot a}=
	\frac 12M_{\rm eff}{\bf\dot u}^T{\bf\dot u}=
	\frac 12M_{\rm eff}\alpha^2{\bf\dot a}^T
	{\bf B}{\bf B}^T{\bf\dot a}.
	\label{Ek2}
\end{equation}

The model matrix
\begin{equation}
    {\bf B}=\sqrt{\frac{5}{4\pi}} 
    \ifx\endpmatrix\undefined\pmatrix{\else\begin{pmatrix}\fi {{3\,
 \varphi+2}\over{4\,\varphi+3}}&-{{3\,\varphi+2}\over{4\,\varphi+3}}&
 0&-{{1}\over{\varphi+2}}&0&{{1}\over{\varphi+2}}\cr -{{\varphi+1
 }\over{4\,\varphi+3}}&{{\varphi+1}\over{4\,\varphi+3}}&0&-{{\varphi+
 1}\over{\varphi+2}}&0&{{\varphi+1}\over{\varphi+2}}\cr -{{2\,\varphi
 +1}\over{4\,\varphi+3}}&-{{2\,\varphi+1}\over{4\,\varphi+3}}&-{{2\,
 \varphi+1}\over{4\,\varphi+3}}&{{\varphi}\over{\varphi+2}}&{{\varphi
 }\over{\varphi+2}}&{{\varphi}\over{\varphi+2}}\cr {{\varphi+1}\over{
 \sqrt{3}\,\left(4\,\varphi+3\right)}}&{{\varphi+1}\over{\sqrt{3}\,
 \left(4\,\varphi+3\right)}}&-{{2\,\left(\varphi+1\right)}\over{
 \sqrt{3}\,\left(4\,\varphi+3\right)}}&-{{\varphi+1}\over{\sqrt{3}\,
 \left(\varphi+2\right)}}&{{2\,\left(\varphi+1\right)}\over{\sqrt{3}
 \,\left(\varphi+2\right)}}&-{{\varphi+1}\over{\sqrt{3}\,\left(
 \varphi+2\right)}}\cr -{{3\,\varphi+2}\over{\sqrt{3}\,\left(4\,
 \varphi+3\right)}}&-{{3\,\varphi+2}\over{\sqrt{3}\,\left(4\,\varphi+
 3\right)}}&{{2\,\left(3\,\varphi+2\right)}\over{\sqrt{3}\,\left(4\,
 \varphi+3\right)}}&-{{1}\over{\sqrt{3}\,\left(\varphi+2\right)}}&{{2
 }\over{\sqrt{3}\,\left(\varphi+2\right)}}&-{{1}\over{\sqrt{3}\,
 \left(\varphi+2\right)}}\cr 
 \ifx\endpmatrix\undefined}\else\end{pmatrix}\fi
\end{equation}

has special properties obtained, or directly from the matrix $\bm B$
of with the help of spherical harmonics sum rules

\begin{equation}
	{\bf B}{\bf B}^T=\frac{3}{2\pi}{\bf I}
	\qquad
	{\bf B}^T{\bf B}=\frac{3}{2\pi}\left({\bf I}-
	\frac{1}{6}{\bf 1}\right)=\frac{3}{2\pi}\bm\Gamma
	\qquad
	{\bf B1}={\bf 0}.
	\label{Bprop}
\end{equation}

Furthermore, the Moore-Pensore pseudo inverse of 
$\bf B$, $\bf B^+$ is

\begin{equation}
    {\bf B}^+={\bf B}^T({\bf BB}^T)^{-1}=
    \frac{2\pi}{3}{\bf B}^T
    \label{Bplus}
\end{equation}

so that

\begin{equation}
    {\bf B}^+{\bf B}={\bf\Gamma}.
\end{equation}

From Eq.(\ref{Ek2}) we have

\begin{equation}
	\frac 12M_S{\bf\dot a}^T{\bf\dot a}=
	\frac 12M_{\rm eff}\frac{3\alpha^2}{2\pi}
	{\bf\dot a}^T{\bf\dot a}
\end{equation}

then

\begin{equation}
	M_{\rm eff}=\frac{2\pi}{3\alpha^2}M_S=\frac 56M_{\rm eq}=284\,\rm kg.
\end{equation}

\section{Movement equation in terms of surface deformation}
\label{moveqsd}
We have seen that the movement equation for the modes of a bare sphere under the action of
$N$ external forces of the type 
$\bm f_a=f_a\delta(\bm x-\bm x_a)\bm e_a$ 
at the positions $\bm x_a$ is given by
\begin{equation}
	{\bf\ddot a} + 2\beta{\bf\dot a}+
	w_0^2{\bf a}=\frac{1}{M_S}\alpha{\bf B}{\bf f}.
\end{equation}

Multiplying this equation by $\alpha{\bf B}^T$ we obtain it in terms of the sphere surface deformation $\bm u$

\begin{equation}
	{\bf\ddot u} + 2\beta{\bf\dot u}+
	w_0^2{\bf u}=\frac{1}{M_S}\alpha^2{\bf B}^T{\bf Bf}
\end{equation}

and using Eq.(\ref{Bprop}) results in

\begin{equation}
	{\bf\ddot u} + 2\beta{\bf\dot u}+
	w_0^2{\bf u}=\frac{3\alpha^2}{2\pi M_S}{\bf\Gamma f}=
	\frac{1}{M_{\rm eff}}{\bf\Gamma f}.
\end{equation}

Finally the movement equation for the deformation of the sphere surface at the position of transducers is

\begin{equation}
	M_{\rm eff}{\bf\ddot u} + 
	2M_{\rm eff}\beta{\bf\dot u}+M_{\rm eff}w_0^2
	{\bf u}={\bf\Gamma}{\bf f}.
\end{equation}

\section{Real vector spherical harmonics}\label{vsharmonics}
The orthogonal real vector spherical harmonics are given by \cite{Landau04}
\begin{equation}
    \boldsymbol{Y}_{\ell m}^{L}(\theta,\phi)=Y_{\ell m}^{\cal R}(\theta,\phi)\bm{\hat r}
    \label{YL}
\end{equation}
\begin{equation}
    \boldsymbol{Y}_{\ell m}^{E}(\theta,\phi)=
        \frac{1}{\sqrt{\ell(\ell+1)}}r\boldsymbol{\nabla}Y_{\ell m}^{\cal R}(\theta,\phi)
    \label{YE}
\end{equation}
\begin{equation}
    \boldsymbol{Y}_{\ell m}^{M}(\theta,\phi)=
        \bm{\hat r}\times\boldsymbol{Y}_{\ell m}^{E}(\theta,\phi)
    \label{YM},
\end{equation}

where the real spherical harmonics $Y_{\ell m}^{\cal R}$ are given by the real and imaginary 
part of the traditional spherical harmonics, Eq.(\ref{realY}), \cite{Jackson}. The real vector spherical harmonics obey the normalization condition

\begin{equation}
    \int\boldsymbol{Y}_{N}^{A}\cdot\boldsymbol{Y}^B_{N'}
    \sin\theta d\theta d\phi=\delta_{NN'}\delta_{AB}.
\end{equation}

\section{Transformation of $h_{ij}$ from wave frame to lab frame}
\label{hWFtohLF}
The polarization tensor for a GW 
propagating in $Z$ direction of the wave frame with polarizations $h_+$ and $h_\times$ is \cite{Merkowitz97}

\begin{equation}
    {\bf h}_{WF}=
    \begin{pmatrix}
        h_+ & h_\times & 0\\
        h_\times & -h_+ & 0\\
        0 & 0 & 0
    \end{pmatrix}.
\end{equation}

Let the matrix ${\bf A}(\theta,\phi,\psi)={\bf R}_z(\psi){\bf R}_y(\theta){\bf R}_z(\phi)$ 
rotates the lab reference frame to the direction $(\theta,\phi,\psi)$
using Euler's-y convention. Any vector $\bm v$ can be rotated by this direction
using the transpose of this matrix 
${\bf A}^T(\theta,\phi,\psi)={\bf R}^T_z(\phi){\bf R}^T_y(\theta){\bf R}^T_z(\psi)$.
In the case GW we are not interested in $\psi$ rotation because this only
mixes the $h_+$, $h_\times$ polarizations. Without the $\psi$ rotation the matrix
${\bf A}$ becomes

\begin{equation}
    {\bf A}=\begin{pmatrix} 
    \cos\theta\cos\phi & \cos\theta\sin\phi & -\sin\theta \cr 
    -\sin\phi           & \cos\phi            & 0 \cr
    \sin\theta\cos\phi & \sin\theta\sin\phi& \cos\theta 
 \end{pmatrix}.
\end{equation}

If we have an incoming wave in the direction of the $z$ axis of the lab frame,
after a rotation to the direction $(\theta,\phi)$ it is seen from the lab frame as

\begin{equation}
    {\bf h}_{LF}={\bf A}^T{\bf h}_{WF}{\bf A}.
\end{equation}

\begin{acknowledgments}
\noindent 

The authors would like to acknowledge \textit{Funda\c{c}\~ao de Amparo \`a Pesquisa do Estado de S\~ao Paulo} (FAPESP) for financial support under the grant numbers 1998/13468-9, 2006/56041-3, 2013/26258-4,
2017/05660-0, 2018/02026-0, and 2020/05238-9. ODA thanks the Brazilian Ministry of Science, Technology and Inovations and the Brazilian Space Agency as well.
Support from the \textit{Conselho Nacional de Desenvolvimento Cientifíco e Tecnológico} (CNPq) is also acknowledged under the grants number 302841/2017-2, 310087/2021-0, and
312454/2021.

\end{acknowledgments}

\bibliographystyle{apsrev4-1}
\bibliography{IOPLaTex}
\end{document}